\documentclass[9pt,twocolumn,twoside]{opticajnl}
\journal{opticajournal} 

\setboolean{shortarticle}{true}

\usepackage{lineno}

\title{Quantum State Tomography in a Third-Order Integrated Optical Parametric Oscillator}

\author[1,*]{Roger Alfredo Kögler}
\author[1]{Gabriel Couto Rickli}
\author[2]{Renato Ribeiro Domeneguetti}
\author[3]{Xingchen Ji}
\author[3,4]{Alexander L. Gaeta}
\author[3,4]{Michal Lipson}
\author[1]{Marcelo Martinelli}
\author[1]{Paulo Nussenzveig}

\affil[1]{Instituto de F\'isica, Universidade de S\~ao Paulo, Caixa Postal 66318, 05315-970 S\~ao Paulo, S\~ao Paulo, Brazil.}
\affil[2]{Center for Macroscopic Quantum States bigQ, Department of Physics, Technical University of Denmark, Fysikvej 307, DK-2800 Kgs. Lyngby, Denmark.}
\affil[3]{Department of Electrical Engineering, Columbia University, New York, New York 10027, USA.}
\affil[4]{Department of Applied Physics and Applied Mathematics, Columbia University, New York, New York 10027, USA.}

\affil[*]{Corresponding author: rogerkogler@alumni.usp.br}

\begin{abstract}
We measured the covariance matrix of the fields generated in an integrated third-order optical parametric oscillator operating above threshold. We observed up to $(2.3 \pm 0.3)$ dB of squeezing in amplitude difference, inferred $(4.9 \pm 0.7)$ dB of on-chip squeezing, while an excess of noise for the sum of conjugated quadratures hinders the entanglement. The degradation of amplitude correlations and state purity for the increasing of the pump power is consistent with the observed growth of the phase noise of the fields, showing {the necessity of strategies for phase noise control aiming at entanglement generation in these systems.}
\end{abstract}

\setboolean{displaycopyright}{false} 

\begin{document}

\maketitle

\section{Introduction}

Novel photonic quantum technologies rely on integrated sources of nonclassical light, generating states that range from single photons to entangled states of bright fields. Optical parametric oscillators (OPOs) are widely employed for this purpose. The development of nanophotonics brought these devices into the microscale domain \cite{kippenberg2004kerr}. Nowadays, they represent a reliable source of entangled photons \cite{yin2021frequency}, being a building block to the realization of integrated quantum information protocols \cite{llewellyn2020chip}. In the continuous variable domain, several important milestones were achieved, such as on-chip optical squeezing using second- ($\chi^{(2)}$) \cite{mondain2019chip,chen2022ultra} and third-order ($\chi^{(3)}$) nonlinearities \cite{dutt2015chip,dutt2016tunable,vaidya2020broadband,zhao2020near,zhang2021squeezed,jahanbozorgi2023generation}. In particular, silicon photonics 
are of great interest due to its compatibility with the CMOS (complementary metal–oxide–semiconductor) fabrication process, enabling seamless integration of photonics and microelectronics in the same chip. Leveraged by its mature manufacturing industry, low losses waveguides are routinely fabricated, resulting in ultra-high quality factor optical micro-cavities \cite{xuan2016high}. 

{Here, we present the full quantum tomography of the complete Gaussian states generated in an on-chip OPO for the first time. Aiming the observation of entanglement in those systems, theoretically predicted in references \cite{gonzalez2017third,bensemhoun2024multipartite}, we reconstruct the four-mode covariance matrix of the output states with a resonator-assisted measurement technique \cite{villar2008conversion,barbosa2013quantum}. Our results reveal unexpected effects resulting from the system dynamics in the studied operation regime.} The present article is organized as follows: in section \ref{sec:methods}, we describe our experimental system and the data analysis methods to reconstruct the covariance matrix from our measurements. Section \ref{sec:results} shows the properties of the different states generated under different pump powers. Finally, we discuss the results and the limits in the production of quantum correlations in Section \ref{sec:conclusion}.

\section{Materials and Methods}
\label{sec:methods}

Our OPO consists of an on-chip silicon nitride microresonator on a silicon oxide substrate. Resonators with high-quality factors and strong light confinement boost intracavity powers and enhance third-order nonlinear interactions between
resonant frequencies and the medium. {The most relevant interactions in our system are self- and cross-phase modulations and the four-wave mixing (FWM), with the last being responsible for populating signal and idler modes. In the process, two photons of the pump mode are annihilated, and signal and idler photons are simultaneously generated, respecting energy conservation. Phase matching condition, necessary for parametric gain around the optical pump, are guaranteed by anomalous group-velocity dispersion, which can be achieved by the combination of the material dispersion and our waveguides geometry (with $2630 \times 730$ nm$^2$ cross-section) \cite{turner2006tailored}. Pairwise photon generation implies in intensity and amplitude correlation. Energy and momentum conservation leads to phase anti-correlation \cite{matsko2005optical,chembo2016quantum,gonzalez2017third,bensemhoun2024multipartite}.}

The micro-cavity is built with a 
closed loop 
resonator with a free spectral range of $80$ GHz, 
separated by a 250 nm gap from the 
bus waveguide. Its loaded and intrinsic quality factors are $Q_L=2$ million and $Q_I=16$ million, respectively. Therefore the resonator is overcoupled for efficient intracavity light extraction, and up to $9.0$ dB of squeezing is expected to be generated for operation 
slightly above the optical threshold \cite{chembo2016quantum}.

\begin{figure}[htbp]
\centering
\includegraphics[width=.45\textwidth]{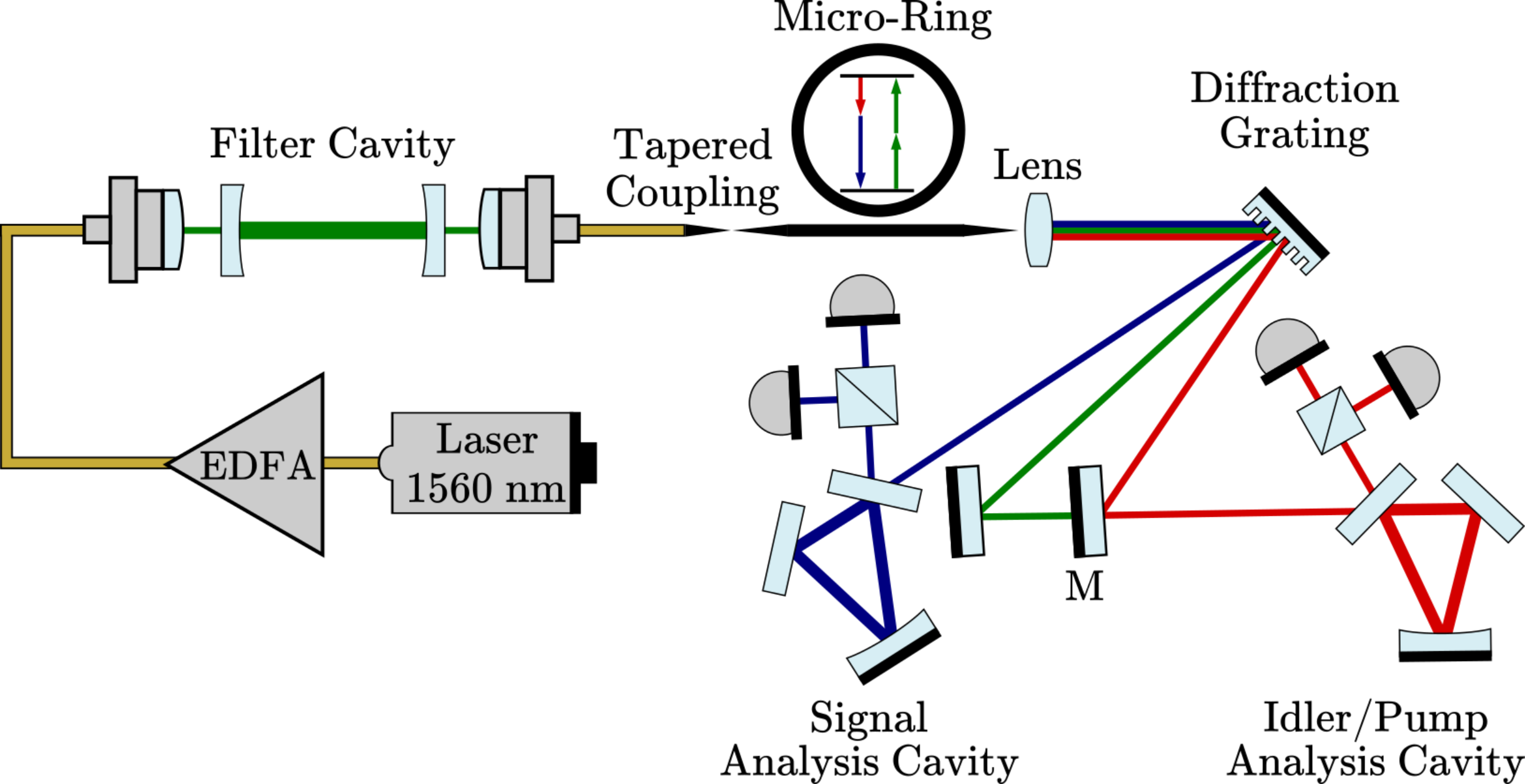}
\caption{\label{setup} Simplified scheme of the experimental setup. 
}
\end{figure}

As schematically illustrated in figure \ref{setup}, 
light from a $1560$ nm diode laser (RIO ORION${}^{\textrm{TM}}$) is amplified by an EDFA (Erbium-doped fiber amplifier). The beam is sent through a filter cavity that reduces the noise close to a coherent state, as detailed in section S2.A of the supplementary material. 
The resulting pump field is coupled into the chip using a tapered fiber, and a coupling efficiency of $70 \%$ is reached with the assistance of inverse tapering design \cite{cardenas2014high}. The guided field is evanescently coupled to the microresonator from the single bus waveguide, in an add-through configuration \cite{yariv2000universal}. We tune the cavity into resonance by thermo-optical effect \cite{zanatta2013thermo} using an integrated micro-heater located over the resonator. Above the oscillation threshold of $\sim 13.0$ mW, bright signal {($1544$ nm)} and idler {($1578$ nm)} fields are produced, with output in the range of a few mW. The output fields leave the chip and are collimated with an objective lens and spatially separated with a diffraction grating ($600$ grooves/mm and $13\%$ losses).

After the separation, signal and idler are reflected by individual analysis cavities, and the photocurrent generated by PIN photodiodes are further processed by a demodulation chain and registered on a computer for further analysis. Removal of mirror M allows the analysis of the pump field, and the use of a pair of detector allows for a permanent verification of the corresponding shot noise level by subtraction of the registered photocurrents.
In this resonator assisted detection scheme, we use the fact that the spectral analysis of the photocurrent $\hat I(t)=\int \hat I_\Omega e^{-i\Omega t}dt$ reveals, on the beatnote of combination of sidebands with the intense carrier field, the measurement of quadratures of these sidebands. Cavities will manipulate the phases of the involved fields, giving access to distinct combination of quadratures that are projected into the amplitude fluctuations of the reflected beam 
\cite{villar2008conversion}. 
Thus we can reconstruct all the elements of the covariance matrix of the states \cite{barbosa2013quantum}. We review the resonator assisted detection scheme and the reconstruction of the single- and two-mode covariance matrices in the section S1 of the supplementary material.

The observables of the quadratures are related to the creation and annihilation operators of the upper and lower sideband modes: $\hat p_{\pm\Omega}=\hat a^\dag_{\pm\Omega}+\hat a_{\pm\Omega}$ and $\hat q_{\pm\Omega}=i(\hat a^\dag_{\pm\Omega}-\hat a_{\pm\Omega})$. These can be associated respectively with the amplitude and phase quadratures of the intense field.
On the other hand, detection process involving the spectral components {($\Omega$)} of the detectors' photocurrent, will necessarily bring the information of the beat of the upper and lower sidebands with the intense mean field \cite{barbosa2013quantum}. Thus the description of the measurements can be conveniently expressed using a basis involving both symmetric and anti-symmetric combinations of these sidebands, defined as $\hat{p}_{\mathfrak{s}} =  ({\hat{p}_{\Omega} + \hat{p}_{-\Omega}})/\sqrt{2}$, $\hat{p}_{\mathfrak{a}} =  ({\hat{p}_{\Omega} - \hat{p}_{-\Omega}})/\sqrt{2}$, $\hat{q}_{\mathfrak{s}} =  ({\hat{q}_{\Omega} + \hat{q}_{-\Omega}})/\sqrt{2}$ and $\hat{q}_{\mathfrak{a}} =  ({\hat{q}_{\Omega} - \hat{q}_{-\Omega}})/\sqrt{2}$.


Therefore the complete state of the Gaussian field involving the pair of sidebands of the converted fields can be associated to the covariance matrix \cite{barbosa2013quantum}
\begin{equation}
\label{eq_covmatrixred}
    \mathbb{V}^{}_{} = \begin{bmatrix}
    \mathbb{V}_{\mathfrak{s}}^{} & \mathbb{C}_{\mathfrak{(s,a)}}^{} \\
    \mathbb{C}_{\mathfrak{(s,a)}}^{\textrm{T}} & \mathbb{V}_{\mathfrak{a}}^{}.
\end{bmatrix}
\end{equation}
The sub-matrices are related to the sideband quadratures through $ \mathbb{V}^{}_{{\mathfrak{s}}} = \frac{1}{2} \left\langle{\mathbf{x}_{{\mathfrak{s}}}^{} \cdot \mathbf{x}_{{\mathfrak{s}}}^{\textrm{T}} + \left( \mathbf{x}_{{\mathfrak{s}}} \cdot \mathbf{x}_{{\mathfrak{s}}}^{\textrm{T}} \right)^{\textrm{T}}}\right\rangle$ and $\mathbb{V}^{}_{{\mathfrak{a}}} = \frac{1}{2} \left\langle{\mathbf{x}_{{\mathfrak{a}}}^{} \cdot \mathbf{x}_{{\mathfrak{a}}}^{\textrm{T}} + \left( \mathbf{x}_{{\mathfrak{a}}} \cdot \mathbf{x}_{{\mathfrak{a}}}^{\textrm{T}} \right)^{\textrm{T}}}\right\rangle$, where we ordered the quadrature operators as $\mathbf{{x}}_{j} = \left[\hat{p}_{j}^{(s)}, \hat{q}_{j}^{(s)},\hat{p}_{j}^{(i)},\hat{q}_{j}^{(i)}\right]^{\textrm{T}}$, $j=\{\mathfrak{s},\mathfrak{a}\}$. The cross-correlation matrix is given by $\mathbb{C}^{}_{({\mathfrak{s}},{\mathfrak{a}})} = \left\langle {\mathbf{x}_{{\mathfrak{s}}}^{} \cdot \mathbf{x}_{{\mathfrak{a}}}^{\textrm{T}} } \right\rangle$. Since spectral component $\hat I(\Omega)\propto  e^{-i\varphi}\hat a_{\Omega}+e^{i\varphi}\hat a^\dag_{-\Omega}$ explicitly brings the contribution of the sidebands, with a phase $\varphi$ of the carrier, the measurement of the noise power $\Delta^2 I^{(m)}=\langle \hat I^{(m)}(\Omega)\hat I^{(m)}(-\Omega)\rangle$ for signal and idler modes ($m=\{s,i\}$), and the cross correlation of the quadratures of the photocurrent  $Re\{\langle \hat I^{(s)}(\Omega)\hat I^{(i)}(-\Omega)\rangle\}$ and $Im\{\langle \hat I^{(s)}(\Omega) \hat I^{(i)}(-\Omega)\rangle\}$, enable a full reconstruction of the covariance matrix $\mathbb{V}^{}_{}$ by the measurement of the photocurrents while scanning the analysis cavities.

\begin{figure}[htbp]
\centering
\includegraphics[width=.45\textwidth]{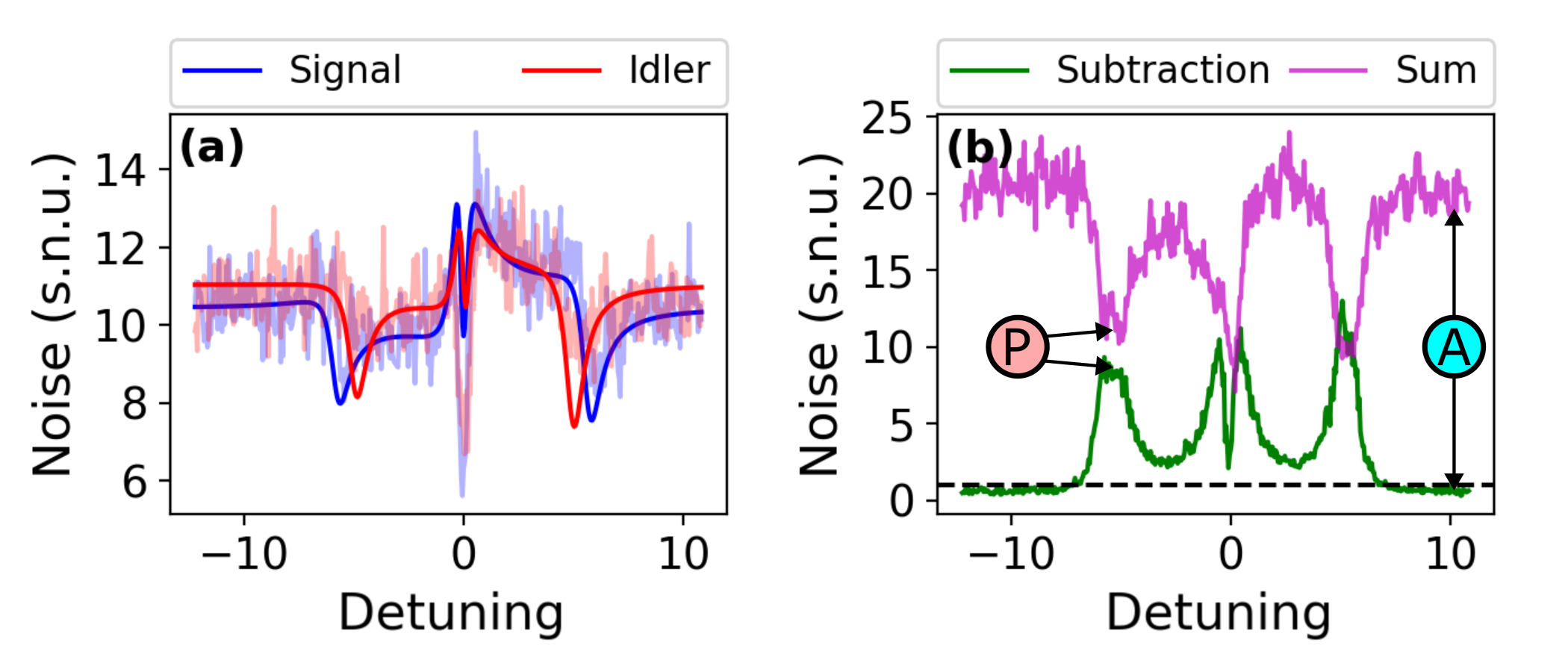}
\caption{\label{fig_PS} \textbf{(a)} Power spectrum of signal and idler fields. Straight lines are the fitted curves of equation ($S10$) of the supplementary material to the faded acquired data in the background. \textbf{(b)} Noise for sum and subtraction of the photocurrents. Two regions of the graphic that give us visual information on the amplitude (A circle) and phase (P circle) are indicated. {s.n.u.: shot noise units.}}
\end{figure}

{Figure \ref{fig_PS} (a) shows one measurement of signal and idler photocurrents noise power at $20$ MHz, where the presented curve corresponds to the state with the optimum value of amplitude difference squeezing. Its covariance matrix elements are given on tables S2 and S3 of the supplemental material. The corresponding frequency was selected in order to avoid excess of technical noise from the system and remain within the detection bandwidth.} All data is corrected by the electronic noise and normalized by the shot noise level. {Analysis cavity detuning are normalized by the idler cavity bandwidth of $4.74$ MHz.} As the cavity is swept around the resonance, one can see the variation on the noise level, indicating unequal quadrature noise. These curves carry part of the information necessary to reconstruct $\mathbb{V}_{\mathfrak{s}}^{}$ and $\mathbb{V}_{\mathfrak{a}}^{}$.

Providing that the measurements were taken synchronously as the cavities sweep around resonance ($\Delta=0$), we can use the sum and subtraction of the photocurrents to verify quantum features in the EPR variables $\hat p_- = (\hat p^{(s)}_s-\hat p^{(i)}_s)/\sqrt{2}$ and $\hat q_+ = (\hat q^{(s)}_s+\hat q^{(i)}_s)/\sqrt{2}$, that can eventually witness entanglement if $\Delta^2 p_-+\Delta^2  q_+<2$ \cite{duan2000inseparability}. 
As shown in figure \ref{fig_PS} (b),  far from resonance, amplitude difference squeezing is observed, as previously evidenced in similar systems operating above threshold \cite{dutt2015chip,dutt2016tunable}, while the sum of the currents demonstrate strong anti-correlation in the amplitudes. On the other hand, strong correlations in phase quadratures are not present. This is an indication that we are generating highly mixed states at the chip output.

\begin{figure}[htbp]
\centering
\includegraphics[width=.45\textwidth]{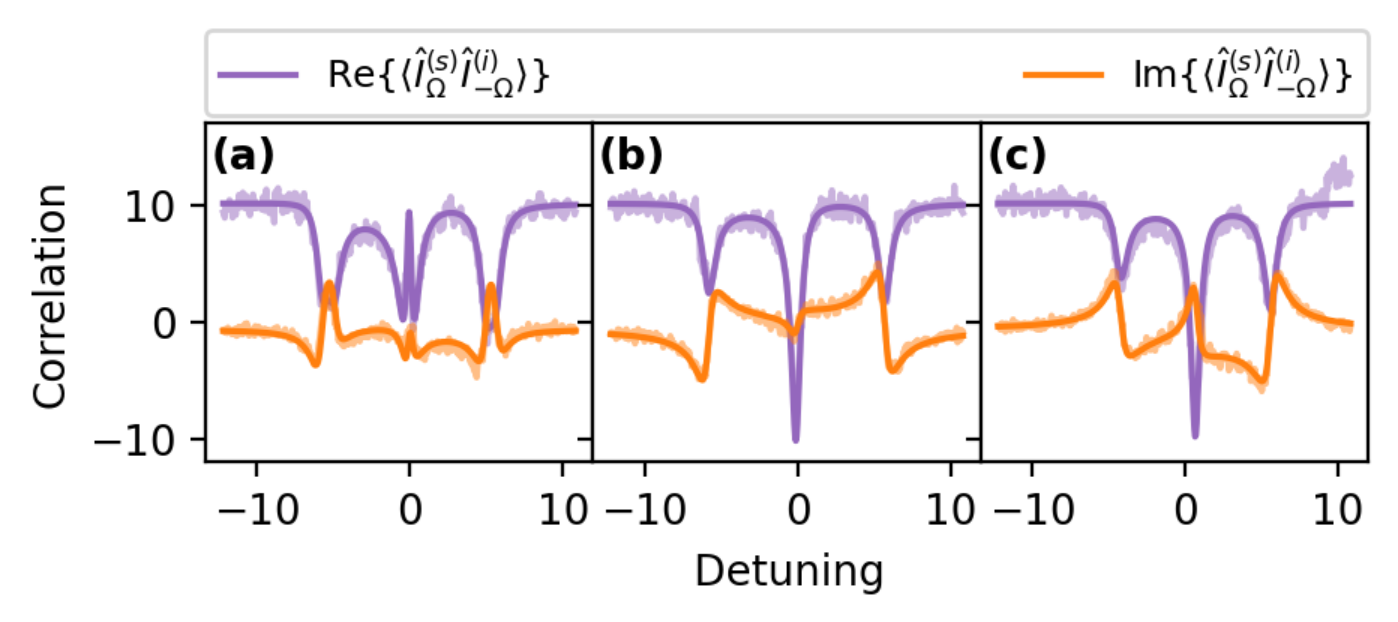}
\caption{\label{fig_correlations} Results from three iterations of the experiment. \textbf{(a)} Simultaneous sweeping of signal and idler analysis cavities, using the same data leading to figure \ref{fig_PS}. \textbf{(b)} Highly detuned idler cavity while sweeping the signal cavity. \textbf{(c)} Highly detuned signal cavity while sweeping the idler cavity. The fitted curves are given by equations ($S15$) and ($S16$) of the supplemental material.}
\end{figure}

Although we retrieve some elements of the covariance matrix from the single- and two-mode power spectra in the sum and subtraction subspaces, the complete reconstruction of the covariance matrix requires access to the cross-correlations in $\mathbb{C}_{\mathfrak{(s,a)}}$. Full tomography of the quantum state is performed through three sequential measurements of the experiment, {which are treated in detail in the section S1 of the supplemental material.} First, the analysis cavities are swept around resonance synchronously, accessing the correlation of on-phase quadratures of the fields. The cross-quadrature correlations are accessed by asynchronous measurements maintaining one cavity far from resonance while sweeping the other. Figure \ref{fig_correlations} shows the data and the fittings of the correlation functions for the same example and conditions presented in figure \ref{fig_PS}.

\section{Results}
\label{sec:results}

Once the measurement procedure and the analysis method for the covariance matrix reconstruction were well defined, we carried out a sequence of experiments varying the pump power, and the final results were evaluated considering the total measurement efficiency of  {$61\%$} ($11\%$ from output coupling into free space, $13\%$ from the diffraction grating, $4\%$ from mismatch with the analysis cavities,  $9\%$ from  optical components, and a detector's quantum efficiency of  $90$\%), giving the on-chip state of the field.

\begin{figure}[ht!]
\centering
\includegraphics[width=.45\textwidth]{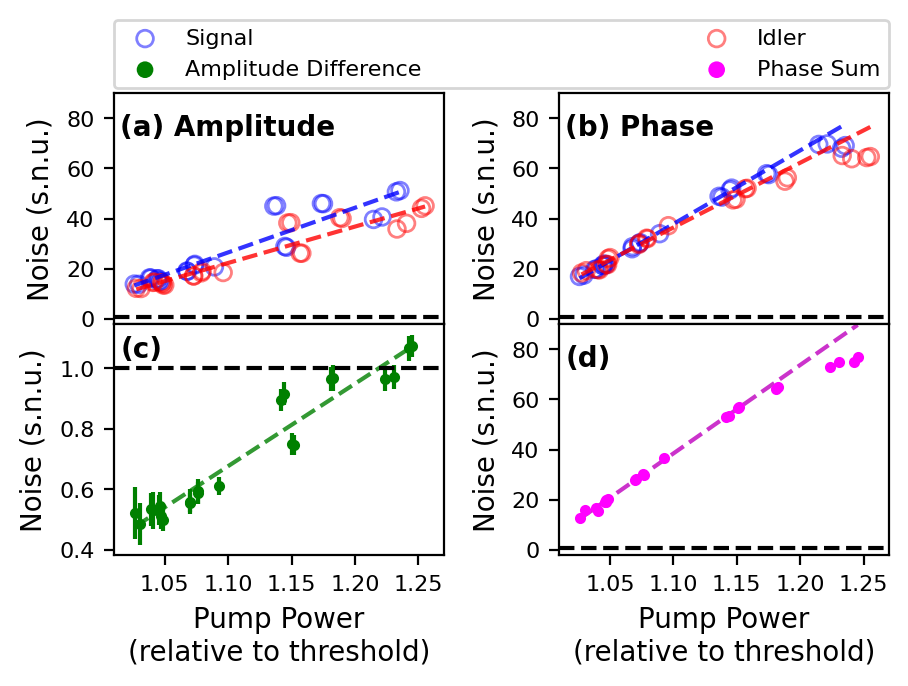}
\caption{\label{fig_quadnoise} Amplitude \textbf{(a)} and phase \textbf{(b)} quadrature noise as a function of the pump power. \textbf{(c)} Amplitude difference squeezing. \textbf{(d)} Phase quadrature sum noise. Dashed lines indicate the shot noise level. Error bars in figures (b)--(d) are buried under the points due to the large scales. {s.n.u.: shot noise units.}}
\end{figure}

Figure \ref{fig_quadnoise} presents the behavior of the quadrature's noise for distinct pump powers (normalized to the $13$ mW threshold).
There is a dramatic increase on the noise power, shown in figures \ref{fig_quadnoise} (a) and (b). As a consequence, any imbalance between the beams caused by the dynamics of the system will lead to the degradation of correlations. This can be seen as a degradation of squeezing in amplitude difference, as shown in figure \ref{fig_quadnoise} (c). {This result is in contrast to what is observed in $\chi^{(2)}$ \cite{jing2006experimental, cesar2009extra} and $\chi^{(3)}$ \cite{guerrero2020quantum} OPOs, where a constant degree of squeezing is observed for an increasing pumping power.} One reason for that is  the mixture of phase and amplitude quadratures due to the phase modulations induced by the dynamics of the third-order OPO \cite{ferrini2014symplectic,gonzalez2017third} and the generation of additional sideband modes \cite{gaeta2019photonic} as the intracavity power builds up. Regarding the sum of the quadratures, correlations are not enough to compensate the excess noise, specially at higher pump powers
(figure \ref{fig_quadnoise} d).

\begin{figure}[htbp]
\centering
\includegraphics[width=.45\textwidth]{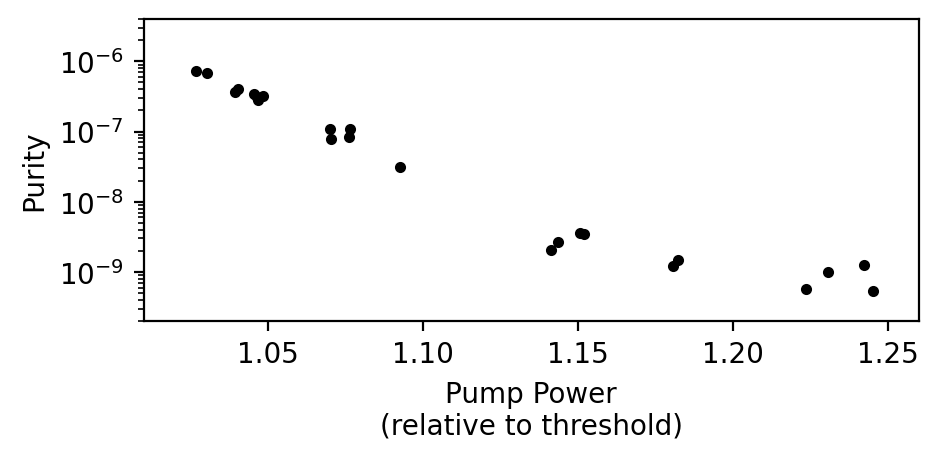}
\caption{\label{fig_purity} Purity of the measured states, {equation ($S25$) of the supplemental material.} Error bars are buried under the points.}
\end{figure}

The purity of the states are readily accessible through the determinant of the covariance matrix \cite{paris2003purity}, {as shown in equation ($S25$) of the supplemental material.} The increasing noise in the quadratures and the degradation of correlations affect the purity of the system, which decreases drastically as we move away from the oscillation threshold, as shown in figure \ref{fig_purity}.
As already stated, no entanglement was expected from the states due to the high phase sum noise. We thoroughly checked this through the application of the PPT criterion \cite{simon2000peres} between all possible partitions of the system (figure S13 of the supplementary material). Intrinsic parametric processes \cite{ferrini2014symplectic,gonzalez2017third} will not degrade the purity of the system, and entanglement should yet be noticed from PPT, in this sense a more efficient method that the direct measurement of the EPR-like quadratures \cite{duan2000inseparability}.

\begin{figure}[htbp]
\centering
\includegraphics[width=.45\textwidth]{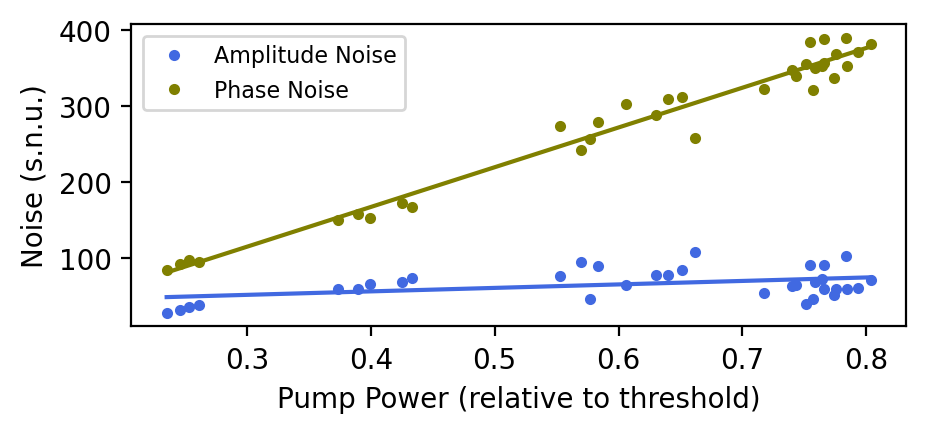}
\caption{\label{fig_pumpnoise} Evolution of amplitude and phase noise of the pump field for the OPO operating below threshold as a function of its power.}
\end{figure}

Therefore, the source of this loss of purity must lie somewhere else. Remaining noise of the pump ($\approx 8 dB$ as seen in Supl. Mat. Sec. 2.A) could not justify the strong noise in the outputs. As a primordial diagnostic of our system, we compute the behavior of the pump noise with the intensity by operating the integrated OPO below the oscillation threshold. We coupled the pump field in the idler analysis cavity to measure its power spectrum. Amplitude and phase noise for increasing values of the pump field are shown in figure \ref{fig_pumpnoise}.
The approximately linear behavior of the fast increasing pump phase noise with pump power is in agreement with previous evidence of photon scattering caused by phonons in $\chi^{(2)}$ crystals \cite{cesar2009extra}. Moreover, previous investigations of the influence of thermal noise in light propagating in waveguides \cite{le2018impact} and microresonators \cite{huang2019thermorefractive} are compatible with a thermorefractive origin. The mitigation of the excessive noise may then be achieved by cooling the system \cite{cernansky2020nanophotonic}. This observation suggests a path for future investigations toward the generation of entangled states. We expect to study the impact of the temperature on the quantum dynamics of the on-chip OPO above threshold.

 
\section{Discussion and Conclusion}
\label{sec:conclusion}

The reconstruction of 
the covariance matrices of intense signal and idler beams generated in an integrated $\chi^{(3)}$ OPO operating above the oscillation threshold is a powerful diagnostic tool to understand the limitations for entangled field generation above the oscillation threshold in these systems.

Using cavity-assisted detection, we were able to perform the tomography of the four-mode state described by the upper and lower sidebands of signal and idler modes. Strong amplitude correlations, beating the standard quantum level, were directly measured and up to $2.3 \pm 0.3$ dB of raw squeezing was observed, {as shown in figure \ref{fig_quadnoise} (c).} Correcting for losses, we infer a total of $4.9 \pm 0.7$ dB of on-chip optical squeezing. This result is against the expected $9.0$ dB of squeezing, given the OPO properties of the study. We attribute this to {unforeseen} on-chip  mechanisms of thermal origin and contamination of the phase noise in amplitude quadrature due to distortions of the noise ellipse induced by Kerr-effect phase modulations \cite{ferrini2014symplectic,gonzalez2017third}.

Observing entanglement remains a challenge, as demonstrated by current results. The four-mode state is highly mixed with a large excess of noise in the phase sum quadrature. Moreover, for stronger pump powers, the noise present in fields’ quadratures increases and amplitude correlations are degraded as the dynamics of the system unbalance signal and idler. 

The noise of the pump, below the oscillation threshold, is consistent with an excess noise of thermal origin observed in\cite{cesar2009extra,le2018impact,huang2019thermorefractive}. {Hence, probing the temperature effects on the integrated OPO is an experimental route that may enable the measurement of entanglement in future work.} Our results shine light on one of the bottlenecks hindering the deterministic generation of entangled states with on-chip silicon based OPOs above threshold.

\begin{backmatter}
\bmsection{Funding} MM and PN acnowledges grant $\#2015/18834-0$, São Paulo Research Foundation (FAPESP). GCR acknowledges grant $\#2021/04829-6$, São Paulo Research Foundation (FAPESP). RRD acknowledges grant $\#2013/26757-0$, São Paulo Research Foundation (FAPESP). This study was financed in part by the Coordenação de Aperfeiçoamento de Pessoal de Nível Superior - Brasil (CAPES) - Finance Code $001$.

\bmsection{Disclosures} The authors declare no conflicts of interest.

\bmsection{Data availability} Data underlying the results presented in this paper are not publicly available at this time but may be obtained from the authors upon reasonable request.

\bmsection{Supplemental document}
See Supplement 1 for supporting content. 

\end{backmatter}

\bibliography{sample}



\ifthenelse{\equal{\journalref}{aop}}{%
\section*{Author Biographies}
\begingroup
\setlength\intextsep{0pt}
\begin{minipage}[t][6.3cm][t]{1.0\textwidth} 
  \begin{wrapfigure}{L}{0.25\textwidth}
    \includegraphics[width=0.25\textwidth]{john_smith.eps}
  \end{wrapfigure}
  \noindent
  {\bfseries John Smith} received his BSc (Mathematics) in 2000 from The University of Maryland. His research interests include lasers and optics.
\end{minipage}
\begin{minipage}{1.0\textwidth}
  \begin{wrapfigure}{L}{0.25\textwidth}
    \includegraphics[width=0.25\textwidth]{alice_smith.eps}
  \end{wrapfigure}
  \noindent
  {\bfseries Alice Smith} also received her BSc (Mathematics) in 2000 from The University of Maryland. Her research interests also include lasers and optics.
\end{minipage}
\endgroup
}{}

\end{document}


\maketitle

\section{Resonator-Assisted Detection}
\label{sec_resonatordetection}

\begin{figure}[ht]
    \centering
    \includegraphics[width=.41\textwidth]{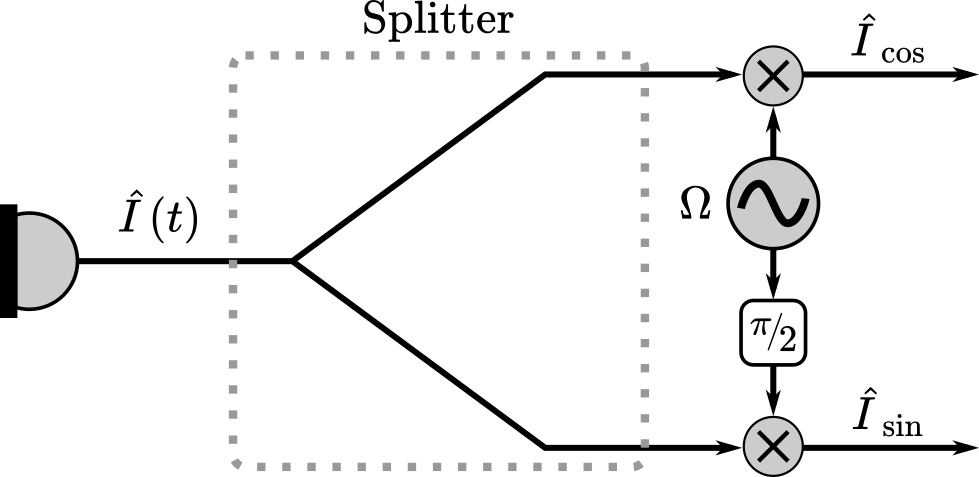}
    \caption{Mixture of the photocurrent with two electronic references of frequency $\Omega$ in quadrature.}
    \label{fig_demod}
\end{figure}

We briefly review the resonator assisted detection method, thoroughly approached in \cite{villar2008conversion,barbosa2013quantum}. Photocurrent operators can be associated to the spectral components, as $\hat I(t)=\int \hat I_\Omega e^{-i\Omega t}dt$. 
In the frequency domain, the non-Hermitian photocurrent operator $\hat{I}_{\Omega} = \hat{I}_{\cos} + i\hat{I}_{\sin}$ for a specific analysis frequency ($\Omega$) can be reconstructed with a double in-quadrature demodulation (figure \ref{fig_demod}), and the outcomes can be associated to the field observables
\begin{align}
    \label{eq_Icos}
    \hat{I}_{\textrm{cos}} = \cos \theta \hat{p}_{\mathfrak{s}} + \sin \theta  \hat{q}_{\mathfrak{s}}, \\
    \label{eq_Isin}
    \hat{I}_{\textrm{sin}} = \cos \theta \hat{q}_{\mathfrak{a}} + \sin \theta  \hat{p}_{\mathfrak{a}},
\end{align}
describing the quadrature operators in the symmetric ($\mathfrak{s}$) and antisymmetric ($\mathfrak{a}$) basis, associated to the usual amplitude ($\hat{p}_{\Omega}$) and phase ($\hat{q}_{\Omega}$) operators by
\begin{align}
    \label{eq_psa}
    \hat{p}_{\mathfrak{s}} = \frac{\hat{p}_{\Omega} + \hat{p}_{-\Omega}}{\sqrt{2}}, \; \; \; \hat{p}_{\mathfrak{a}} = \frac{\hat{p}_{\Omega} - \hat{p}_{-\Omega}}{\sqrt{2}}, \\
    \label{eq_qsa}
    \hat{q}_{\mathfrak{s}} = \frac{\hat{q}_{\Omega} + \hat{q}_{-\Omega}}{\sqrt{2}},  \; \; \; \hat{q}_{\mathfrak{a}} = \frac{\hat{q}_{\Omega} - \hat{q}_{-\Omega}}{\sqrt{2}},
\end{align}
{where the frequency $\pm\Omega$ is associated to the sideband modes of the intense mean field, considered as a carrier at frequency $\omega$.}

The coefficients of the covariance matrix can be retrieved from the power spectrum of the fields and the expectation values of the cross products of the measured photocurrents. By ordering the canonical commuting operators of signal and idler modes as 
\begin{equation}
     {\hat{x}}_{j} = \left[\hat{p}_{j}^{(s)}, \hat{q}_{j}^{(s)},\hat{p}_{j}^{(i)},\hat{q}_{j}^{(i)}\right]^{\textrm{T}}, \; \; \; \; j=\{\mathfrak{s},\mathfrak{a}\},
\end{equation}
one can write the covariance matrix as \cite{barbosa2013quantum}
\begin{equation}
\label{eq_covmatrixred}
    \mathbb{V}^{}_{} = \begin{bmatrix}
    \mathbb{V}_{\mathfrak{s}}^{} & \mathbb{C}_{\mathfrak{(s,a)}}^{} \\
    \mathbb{C}_{\mathfrak{(s,a)}}^{\textrm{T}} & \mathbb{V}_{\mathfrak{a}}^{}
\end{bmatrix}.
\end{equation}
The main diagonal matrices are respectively related to purely symmetric and antisymmetric correlations as $ \mathbb{V}^{}_{{\mathfrak{s}}} = \frac{1}{2} \left\langle{\mathbf{x}_{{\mathfrak{s}}}^{} \cdot \mathbf{x}_{{\mathfrak{s}}}^{\textrm{T}} + \left( \mathbf{x}_{{\mathfrak{s}}} \cdot \mathbf{x}_{{\mathfrak{s}}}^{\textrm{T}} \right)^{\textrm{T}}}\right\rangle$ and $\mathbb{V}^{}_{{\mathfrak{a}}} = \frac{1}{2} \left\langle{\mathbf{x}_{{\mathfrak{a}}}^{} \cdot \mathbf{x}_{{\mathfrak{a}}}^{\textrm{T}} + \left( \mathbf{x}_{{\mathfrak{a}}} \cdot \mathbf{x}_{{\mathfrak{a}}}^{\textrm{T}} \right)^{\textrm{T}}}\right\rangle$. Explicitly, the matrices assume the form
\begin{align}
\label{eq_covmatsymm}
    \mathbb{V}^{}_{{\mathfrak{s}}} = \begin{bmatrix}
    \alpha^{(s)} & \gamma^{(s)} & \mu & \xi \\
    \gamma^{(s)} & \beta^{(s)} & \zeta & \nu \\
    \mu & \zeta & \alpha^{(i)} & \gamma^{(i)} \\
    \xi & \nu & \gamma^{(i)} & \beta^{(i)}
\end{bmatrix},
\end{align}
\begin{align}
    \mathbb{V}^{}_{{\mathfrak{a}}} = \begin{bmatrix}
    \beta^{(s)} & -\gamma^{(s)} & \nu & -\zeta \\
    -\gamma^{(s)} & \alpha^{(s)} & -\xi & \mu \\
    \nu & -\xi & \beta^{(i)} & -\gamma^{(i)} \\
    -\zeta & \mu & -\gamma^{(i)} & \alpha^{(i)}
\end{bmatrix}.
\end{align}
where the indexes (s) and (i) are respectively indicating signal and idler. Finally, the cross correlation matrix is given by
\begin{align}
\label{eq_covmatcorr}
    \mathbb{C}^{}_{({\mathfrak{s}},{\mathfrak{a}})} = \left\langle {\mathbf{x}_{{\mathfrak{s}}}^{} \cdot \mathbf{x}_{{\mathfrak{a}}}^{\textrm{T}} } \right\rangle = \begin{bmatrix}
    \delta^{(s)} & 0 & \kappa & -\eta \\
    0 & \delta^{(s)} & -\tau & -\lambda \\
    -\lambda & \eta & \delta^{(i)} & 0 \\
    -\tau & \kappa & 0 & \delta^{(i)}
\end{bmatrix}.
\end{align}

\begin{figure}[ht]
    \centering
    \includegraphics[width=.6\textwidth]{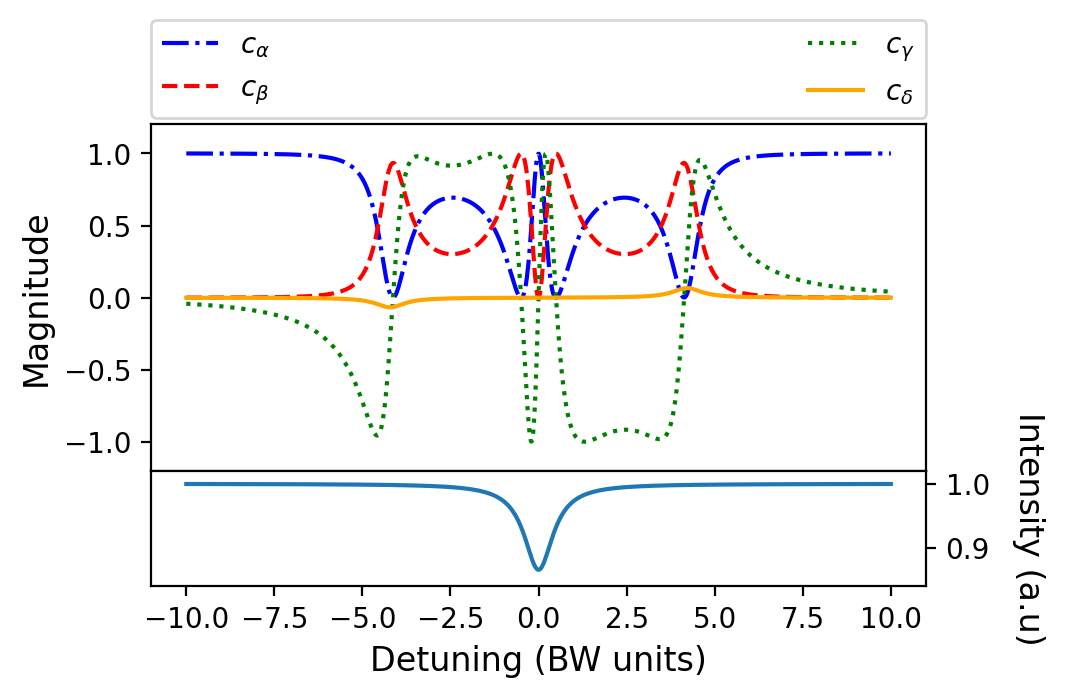}
    \caption{Behavior of the functions $c_j, j=\{\alpha,\beta,\gamma,\delta\}$ as a function of cavity detuning. We use the parameters of the idler cavity in table \ref{table_analysiscavities} as inputs for equations (\ref{eq_gplus})--(\ref{eq_dip}). We also fixed the analysis frequency at $20$ MHz to match the results given in the main text. The bottom curve shows the reflection curve of the analysis cavity.}
    \label{fig_powerspecparams}
\end{figure}

The power spectrum of the output of the analysis cavity, as a function of the cavity detuning, is related to the covariance matrix coefficients as \cite{villar2008conversion,barbosa2013quantum}
\begin{align}
    S(\Delta,\Omega) &= \frac{1}{2}\left\langle{\hat{I}_{\textrm{cos}}^2}\right\rangle + \frac{1}{2}\left\langle{\hat{I}_{\textrm{sin}}^2}\right\rangle \nonumber \\
    &= c_{\alpha} \alpha + c_{\beta} \beta + c_{\gamma} \gamma + c_{\delta} \delta + \Delta^2 \hat{v},
\label{eq_spectraldensitycavity}
\end{align}
where vacuum term couples into the cavity by general losses, such as leak of the mirrors,  and is taken as $\Delta^2 \hat{v} = 1 - c_{\alpha} - c_{\beta}$. The functions $c_{\alpha} = \left\lvert{g_+^{}}\right\rvert^2$, $c_{\beta} = \left\lvert{g_-^{}}\right\rvert^2$, $c_{\gamma} = 2 \textrm{Re}\{g_+^* g_-^{} \}$  and $c_{\delta} = 2 \textrm{Im}\{g_+^* g_-^{} \}$ are dependent on the cavity detuning ($\Delta$) and the analysis frequency ($\Omega$). They depend on the parameters of the individual cavities, which are completely modeled by the mirrors reflectance ($R(\Delta,\Omega)$), given by
\begin{align}
\label{eq_gplus}
    g_{+}(\Delta,\Omega) &= \frac{\left(R(\Delta,\Omega) + R(\Delta,-\Omega)\right)}{2}, \\
    g_{-}(\Delta,\Omega) &= \frac{i\left(R(\Delta,\Omega) - R^*(\Delta,-\Omega)\right)}{2},
\end{align}
where
\begin{align}
    R(\Delta,\Omega) = \frac{r^*(\Delta)}{\left\lvert{r(\Delta)}\right\rvert} r\left(\Delta + \frac{\Omega}{\Delta_{\textrm{BW}}^{\textrm{AC}}}\right).
\end{align}
The reflection coefficients, and as a consequence the $g_\pm^{}$ functions, are experimentally determined by the measurement of the depletion magnitude of the reflected field, named dip ($d = \left\lvert{r(0)}\right\rvert^2$), and the analysis cavity bandwidth ($\Delta_{\textrm{BW}}^{\textrm{AC}}$). For high finesse cavities, the reflection coefficient is
\begin{equation}
\label{eq_dip}
    r(\Delta) = - \frac{\sqrt{d} - 2i\Delta}{1-2i\Delta}.
\end{equation}
Hence, following the parameters of table \ref{table_analysiscavities} we can verify the accessibility of each parameter according to the detuning of the cavities. The continuous measuring of the photocurrents while the cavities sweep around the resonance peak allows us to measure all the power spectrum coefficients, as they contribute differently to the function (\ref{eq_spectraldensitycavity}) depending on the detuning of the cavity. Figure \ref{fig_powerspecparams} show the dependence of the functions $c_j, j=\{\alpha,\beta,\gamma,\delta\}$ with the detuning of the analysis cavity.

{As an example, consider an hypothetical state with amplitude and phase normalized noises respectively given by $\Delta^2 \hat{p} = 1$ and $\Delta^2 \hat{q} = 2$. The spectral power, equation (\ref{eq_spectraldensitycavity}), is shown in figure \ref{fig_ellipserot}. The chosen cavity parameters are equivalent to the signal cavity, table \ref{table_analysiscavities}. Note that as we sweep the cavity, different terms of equation (\ref{eq_spectraldensitycavity}) become relevant, as shown in figure \ref{fig_powerspecparams}. The points referent to amplitude and phase fluctuations on the synchronous detection shown in figure 2 of the main text follow the positions $5$ (equivalent to $1$) and $2$ of the presented example as an indication of the extracted information from the detection method. 
\begin{figure}[ht]
    \centering
    \includegraphics[width=.75\textwidth]{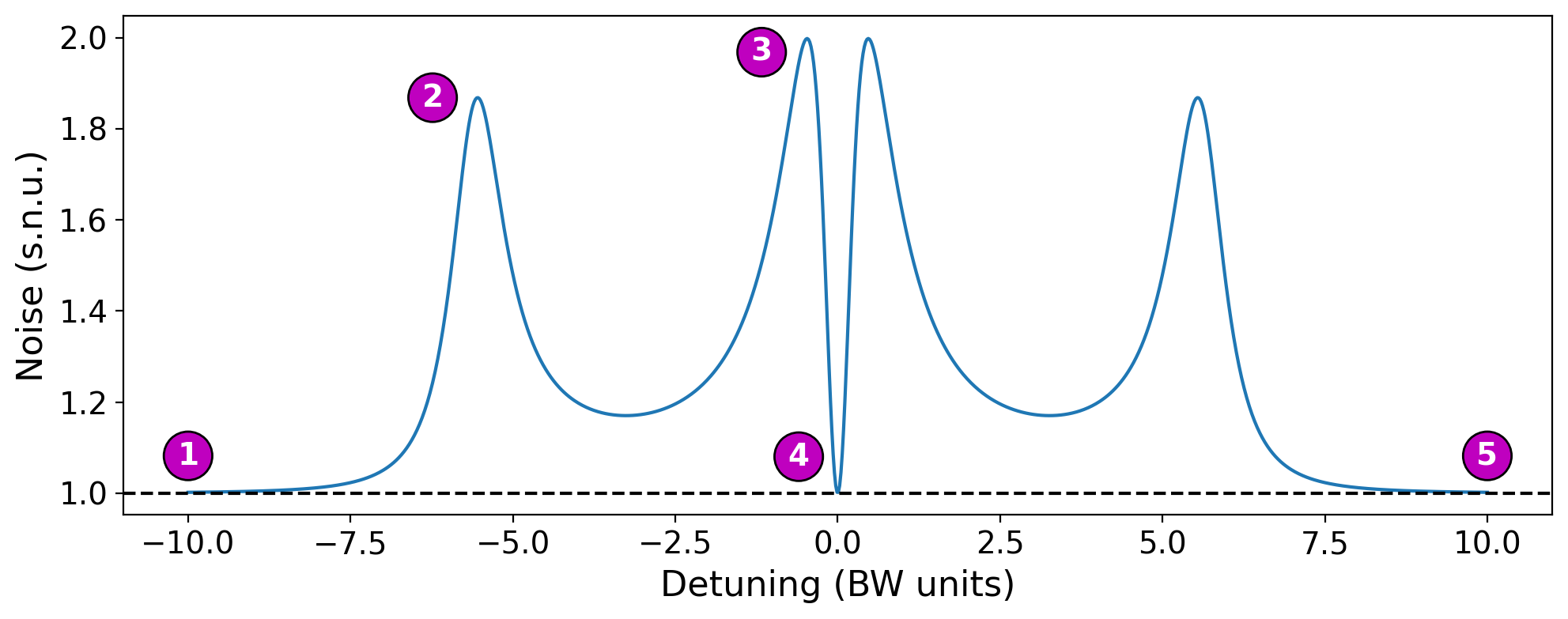}
    \caption{{Spectral density of a thermal field with excess of phase noise. This power spectrum is equivalent to the case where $\alpha = 1$, $\beta = 2$, $\gamma = \delta = 0$. The marked points in the graphic stand to the following: $1$) far from resonance the cavity does not interfere in the noise ellipse and only amplitude fluctuations are measured. $2$) At resonance with the demodulated sideband (in this example $20$ MHz) we have a full phase shift between amplitude and phase noise while leaving the carrier field undisturbed. The small depletion in comparison to the peak $3$ is due to vacuum fluctuations disturbances. 3) This second $\pi/2$ phase-shift is due to the effect of the cavity on the carrier, where now the vacuum only attenuates its the mean field and does not disturb significantly the measured quadrature. 4) At resonance with the carrier, the noise ellipse suffers a $\pi$ phase-shift and the amplitude noise is again accessible. The dashed line represents the shot noise level and s.n.u. stands for shot noise units.}}
    \label{fig_ellipserot}
\end{figure}}

\begin{figure}[ht]
    \centering
    \includegraphics[width=\textwidth]{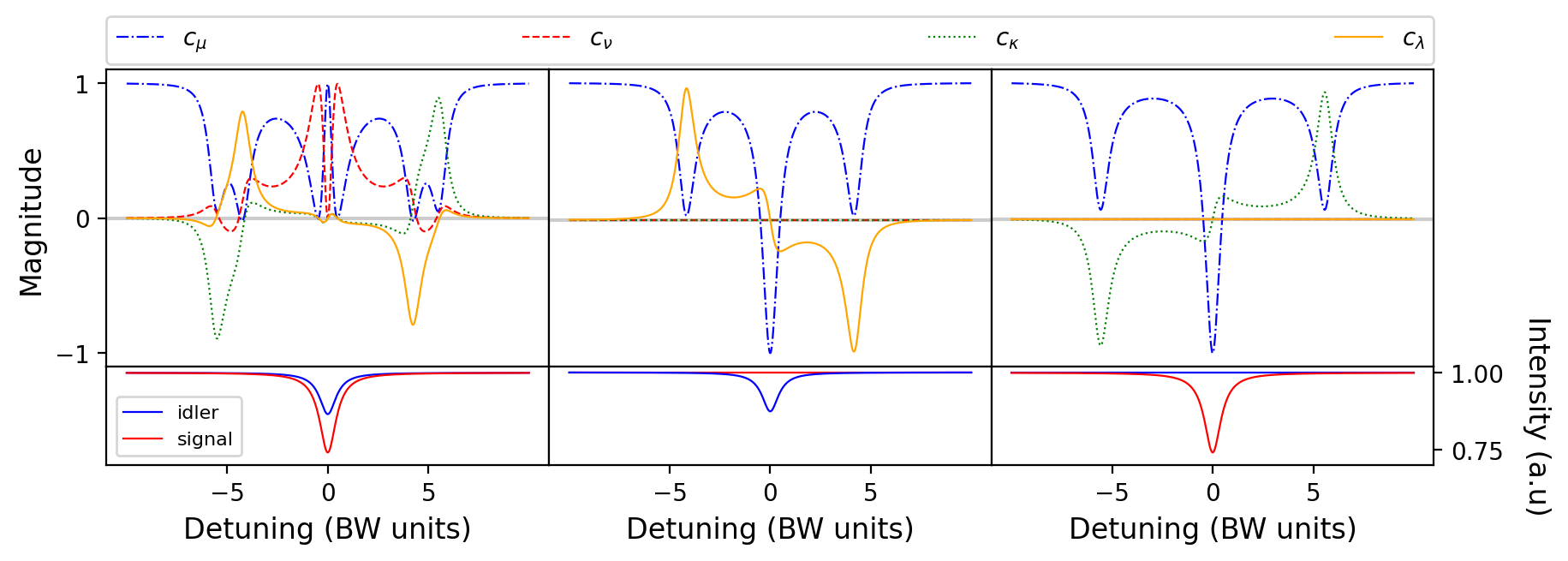}
    \caption{Behavior of the functions $c_j, j=\{\mu,\nu,\kappa,\lambda\}$ as a function of cavity detuning for three different situations. From left to right: synchronous sweeps of the analysis cavities, sweep of the idler cavity while maintaining the signal cavity out of resonance and vice-versa, as indicated in the bottom graphics. We use the parameters of the signal and idler cavities in table \ref{table_analysiscavities} as inputs for equations (\ref{eq_gplus})--(\ref{eq_dip}). Here, we also fixed the analysis frequency at $20$ MHz to match the results given in the main text.}
    \label{fig_rotationmutolambda}
\end{figure}

\begin{figure}[ht]
    \centering
    \includegraphics[width=\textwidth]{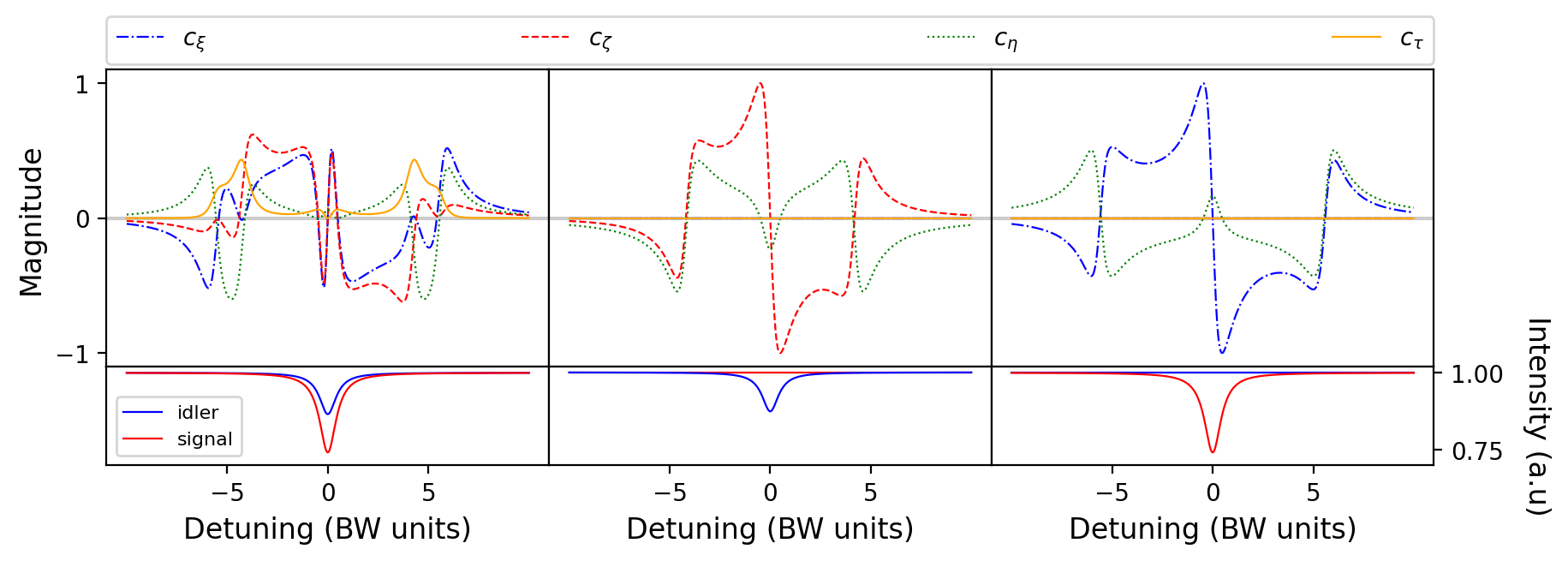}
    \caption{Same as figure \ref{fig_rotationmutolambda}, but for remaining functions $c_j, j=\{\xi,\zeta,\eta,\tau\}$.}
    \label{fig_rotationxitotau}
\end{figure}

The other parameters are retrieved from the covariance terms of the photocurrents:
\begin{align}
    \label{eq_realcorrcavity}
    \textrm{Re}\left\{ \left\langle{\hat{I}_{\Omega}^{(s)}\hat{I}_{-\Omega}^{(i)}}\right\rangle \right\} &= c_{\mu} \mu + c_{\nu} \nu +  c_{\kappa} \kappa + c_{\lambda} \lambda \nonumber \\
    &+ c_{\xi} \xi + c_{\zeta} \zeta + c_{\eta} \eta + c_{\tau} \tau, \\
    \label{eq_imagcorrcavity}
    \textrm{Im}\left\{ \left\langle{\hat{I}_{\Omega}^{(s)}\hat{I}_{-\Omega}^{(i)}}\right\rangle \right\} &= - c_{\eta} \mu - c_{\tau} \nu +  c_{\xi} \kappa + c_{\zeta} \lambda \nonumber \\ 
    &- c_{\kappa} \xi - c_{\lambda} \zeta + + c_{\mu} \eta + c_{\nu} \tau,
\end{align}
where $g_+^{*(s)} g_+^{(i)} = c_{\mu} + ic_{\eta}$, $g_-^{*(s)} g_+^{(i)} = c_{\zeta} + ic_{\lambda}$, $g_-^{*(s)} g_-^{(i)} = c_{\nu} + ic_{\tau}$, and $g_+^{*(s)} g_-^{(i)} = c_{\xi} + ic_{\kappa}$. As in the previous case, all the functions are related with the analysis cavities detuning relative to the carrier mode and their optical parameters. Figures \ref{fig_rotationmutolambda} and \ref{fig_rotationxitotau} shows the behavior of each function as a function of the cavity detuning for the three different experimental acquisitions. That is, with both analysis cavities sweeping through resonance simultaneously and for the permutation of one cavity sweeping while the other is maintained fixed out of resonance (or, equivalently, maintained exactly in resonance with the carrier). {One should note that different terms become more relevant depending on the experimental acquisition. We retrieve all the covariance matrix parameters by imposing that the fitting parameters of equations ($S15$) and ($S16$) to the three different configurations should be equal.}

\section{Detailed Experimental Setup}

The schematic setup of figure 1 of the main text can be divided in three main parts: the optical pump preparation, the integrated OPO, and the detection scheme. Next, we will approach each part individually.

\subsection{Optical Pump Preparation}

\begin{figure}[ht]
    \centering
    \includegraphics[width=.75\textwidth]{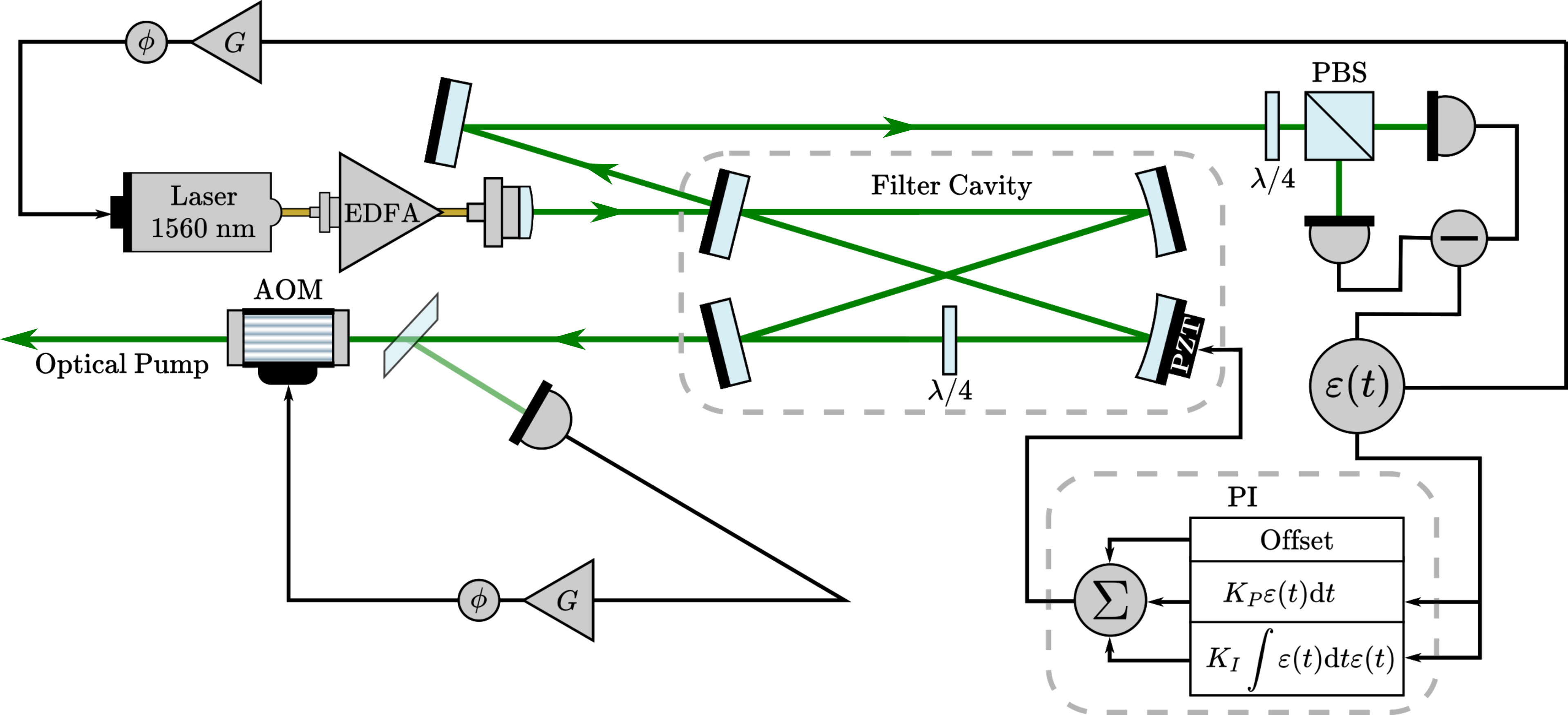}
    \caption{Near coherent optical pump generation scheme. A $1560$ nm diode laser source is amplified by an erbium-doped fiber amplifier (EDFA) and sent to a bow-tie configuration filter cavity. A quarter waveplate was introduced inside the cavity in order to slightly increase the optical losses, which helped on the stabilization of the system. The cavity is locked to the laser frequency with the Hänsch-Couillaud method using a proportional-integral system with the constants $K_P$ and $K_I$ respective to the proportional and the integrator actuators. Current modulations are fed back to the laser in order to mitigate undesired fluctuations. Further cleaning in the cavity output is done with an acousto-optic modulator (AOM). The gain $G$ and phase control $\phi$ necessary for the optimization of the feedback and feedforward systems are represented along the electrical paths.}
    \label{fig_pumpsetup}
\end{figure}

\begin{figure}[ht]
    \centering
    \includegraphics[width=.75\textwidth]{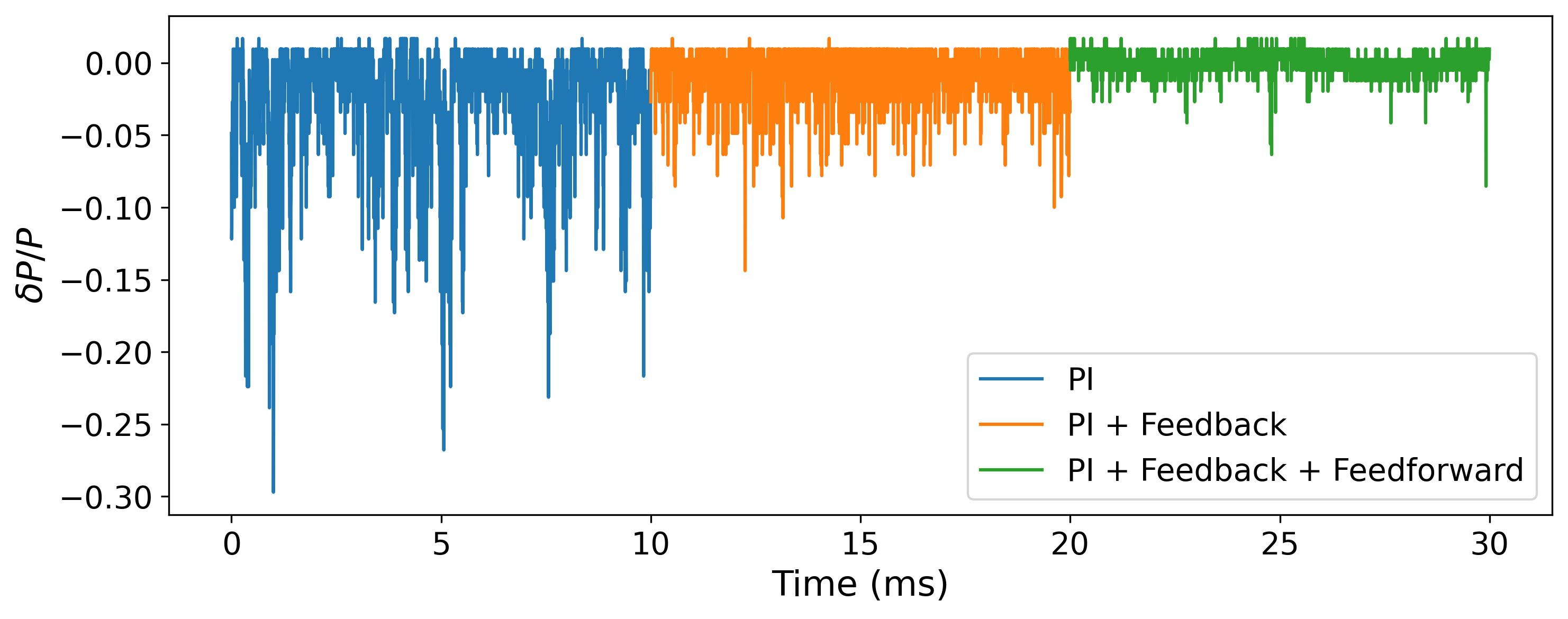}
        \caption{{Evolution of the output pump fluctuations with the actuation of the different stabilization systems. Blue: the PZT alone, feed by the PI system, is unable to mitigate the strong fluctuations. In this configuration we have a relative standard deviation (RSD) of $4.2\%$ and peak-to-peak fluctuations of up to $25\%$ are observed. Orange: a good improvement is seen from the feedback modulations in the laser current, with $\textrm{RSD}=1.6\%$. Green: A cleaner signal, $\textrm{RSD}=0.7\%$, is obtained with the aid of the feed forward power control of the AOM.}}
    \label{fig_pumpfluctu}
\end{figure}

As described in the main text, we use a $1560$ nm RIO ORION${}^{\textrm{TM}}$ diode laser followed by an erbium-doped fiber amplifier as our light source. In order to clean the excessive noise present in this optical beam, we introduce a filtering system, as shown in figure \ref{fig_pumpsetup}.
The filter cavity has an optical path of $4.9$ m, equivalent to a free spectral range of $\Delta_{\textrm{FSR}}=61.2$ MHz, a finesse of $\mathcal{F} = 220.0$ and a bandwidth of $\Delta_{\textrm{BW}} = 278.5$ kHz. The narrow bandwidth of the system contributes to mitigating the excess noise of the sidebands around the central frequency. We lock the cavity in resonance with the Hänsch-Couillaud method \cite{hansch1980laser}, where we feedback the PZT of the moving mirror with a proportional-integral controller. Without additional actuation in our pump, intensity fluctuations at the output of the cavity of the order of $20 \%$ of the optical power were observed in frequencies of hundreds of kHz. {Due to the active stabilization of our OPO, low frequency fluctuations (distant from our measurement band) would jeopardize the OPO stabilization on the range of Hz to kHz.} Since our piezoelectric has a $24$ kHz cutoff, we needed to use other methods to mitigate such fluctuations. First, we used the Hänsch-Couillaud monitoring signal to feedback the current of the laser, thus controlling the intrinsic phase jitter from the diode. Since this was not sufficient to achieve a pump with low intensity fluctuations, we introduced an acousto-optic modulator in a feed forward mechanism to mitigate the undesired fluctuations to an order of $1 \%$ of the optical power. {Figure \ref{fig_pumpfluctu} shows the effect of the combination of our stabilization methods in our pump. The proportional and integration electronic signals fed to the cavity PZT and the gain and phase parameters fed to the laser current and the AOM were optimized monitoring this signal.}

\begin{figure}[ht]
    \centering
    \includegraphics[width=.50\textwidth]{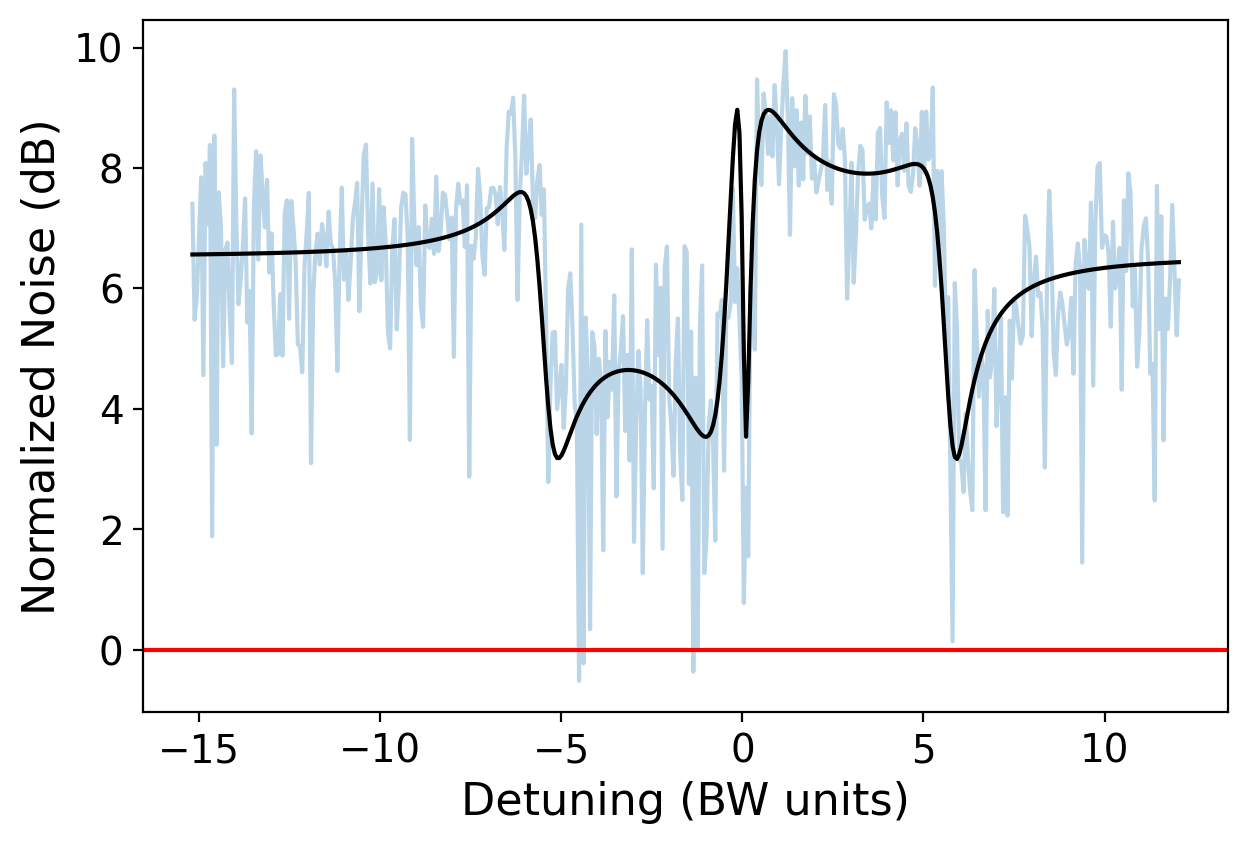}
    \caption{Filtered pump noise as a function of the analysis cavity detuning. The faded blue noise is the experimental data corrected the $13$ mW power and the black solid line is the noise ellipse fitting of equation \ref{eq_spectraldensitycavity}. The red line in $0$ dB is the shot noise reference.}
    \label{fig_filteredpump}
\end{figure}

As a result, we obtained a stable low-noise pump field, where the filter cavity reduced the optical noise by more than $30$ dB for relevant powers. In figure \ref{fig_filteredpump} we show a prediction of the optical pump density spectrum at $13$ mW, which is the oscillation threshold of our OPO. This measurement was made by coupling the filtered pump in the idler analysis cavity, after being coupled to the bus waveguide of the chip. Note that the measurements were taken for highly attenuated pump powers since our detection system is calibrated for the optical powers of signal and idler, that is, hundreds of micro Watts. {Although not coherent, the noise levels of the pump state were drastically reduced while the filter cavity output remained stable. Contributing factors to the pump excess of noise are attributed to remaining unfiltered fluctuations present on the light source and the modulations coming from the AOM \cite{liu2018acousto}.}

\subsection{Chip Stabilization}

\begin{figure}[ht]
    \centering
    \includegraphics[width=.95\textwidth]{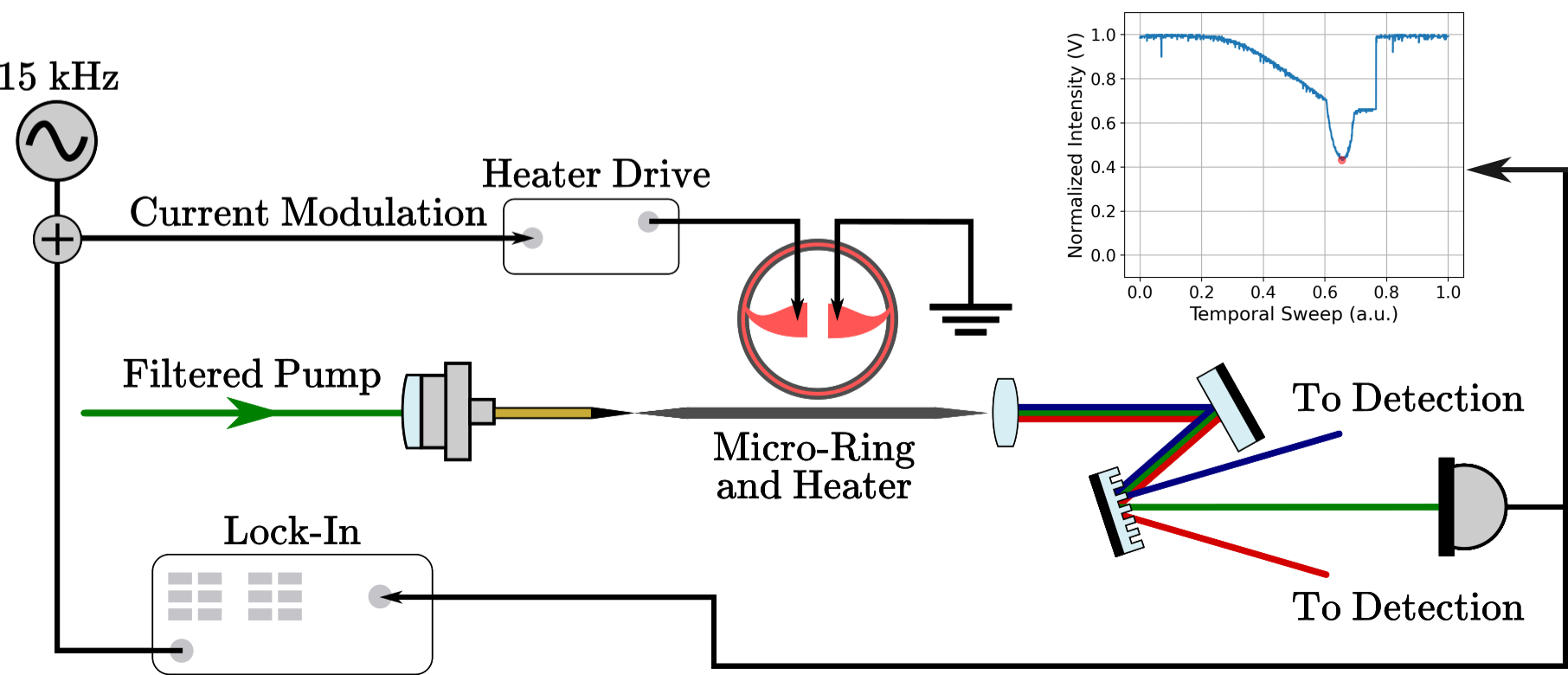}
    \caption{OPO stability system. We modulate the micro-heater at $15$ kHz. The separated pump photocurrent is sent to a lock-in system, added to the $15$ kHz modulation, to keep the resonant condition stable. The inset shows the separated pump response when a voltage ramp is sent to the microresonator to sweep around resonance. The stability point where we lock the system is indicated by the red dot.}
    \label{fig_chiplock}
\end{figure}

Once we have the pump prepared, we need to stabilize the on-chip OPO in resonance, as schematic shown in figure \ref{fig_chiplock}. Thermal drifts, that would eventually take the system out of resonance, were mitigated with a temperature control that maintain the chip close to the room temperature ($\sim 25$ \textdegree C). This is done with a thermo-electric cooler located under the copper basis of the chip, which is actively controlled with a PI system.

The Kerr effect distorts the transmission figure of the sweeping microcavity around resonance with a characteristic bistable signature \cite{almeida2004optical}. Oscillation could be noticed by pump depletion, leading to  an additional dip due to the transfer of energy from the pump to other modes is generated \cite{matsko2005optical}.

Stable signal and idler generation are achieved by actively maintaining the micro-cavity in resonance with the laser using a dither-and-lock system acting on the micro-heater with a modulation of 15 kHz. Without the locking system, mode hopping in the signal and idler frequencies was observed, impairing the coupling with the analysis cavities.

{A pictorial diagram of the fields of interest is presented in figure \ref{fig_sidebands}.
\begin{figure}[ht]
    \centering
    \includegraphics[width=.95\textwidth]{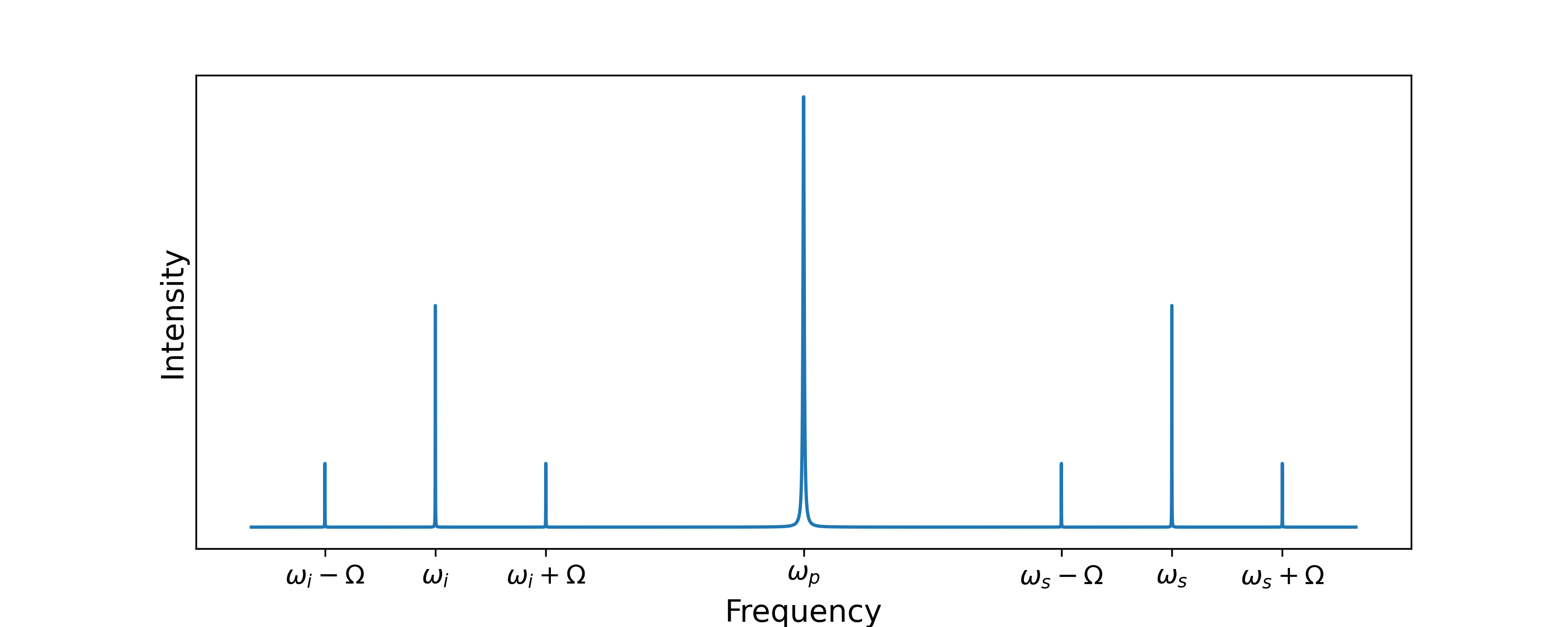}
    \caption{{Frequency diagram ilustrating the central pump, signal and idler fields and their upper and lower sidebands.}}
    \label{fig_sidebands}
\end{figure}
Due to the great difference between absolute values of the relevant frequencies, our representation is not on scale. The frequencies in which the generation of signal and idler are possible are delimited by the OPO bandwidth ($200$ MHz). In principle, signal and idler sidebands within this limit could be investigated, however our detection bandwidth limits the analysis frequency $\Omega$ in $35$ MHz around the carrier fields, beyond which the signal to noise ratio is highly degraded. Specifically, we chose the analysis frequency of $20$ MHz, which is within the OPO bandwidth. This frequency is also relatively far from the carrier frequency, where technical noise from the system is more significant. At last, our demodulation chain bandwidth is of $300$ kHz, which defines the acquisition rate of the system. One should note that the four-mode states reconstructed in our experiment lie in the Hilbert space expanded by the modes $\omega_s-\Omega$, $\omega_s+\Omega$, $\omega_i-\Omega$ and $\omega_i+\Omega$.}


\subsection{Detection System}

At last, the detection system is composed by individual analysis cavities used to scan the signal and idler fields. Relevant parameters of the used resonators are given in table \ref{table_analysiscavities}. The bandwidth and the dip are directly related to the fitting procedures of the main text through equations (\ref{eq_spectraldensitycavity})--(\ref{eq_imagcorrcavity}).
As the access to different terms of the covariance matrix is dependent on the sweeping of the analysis cavities (see figures \ref{fig_powerspecparams}, \ref{fig_rotationmutolambda} and \ref{fig_rotationxitotau}), the scanning velocity of the cavities used in the experiment should also consider the cavity bandwidth, intrinsically connected to the finesse and the FSR by $\mathcal{F} = \Delta_{\textrm{FSR}}/\Delta_{\textrm{BW}}$. Note that the mode matching of the beams to the cavity is given as a mean value since environmental conditions induce variances in the cavities alignment while we operate the system to perform a sequence of measurements.

\begin{table}[htbp]
\centering
\caption{\bf Experimentally determined parameters of the analysis cavities}
\begin{tabular}{ccc}
\hline
\textbf{Parameter}	&	Signal Cavity & Idler Cavity \\
\hline
Free Spectral Range	& $1.03$ GHz & $1.03$ GHz   \\
Bandwidth	& $3.56$ MHz & $4.74$ MHz \\
Finesse	& $290$& $218$ \\
Dip & $25.8\%$ & $13.4\%$ \\
Mode Matching & $97.5\%$ & $95.7\%$\\
\hline
\end{tabular}
  \label{table_analysiscavities}
\end{table}

We use balanced detection schemes to simultaneously access the standard quantum limit and the noise of each beam reflected by the analysis cavities. {This is possible by computing the respective subtraction and sum of the balanced photocurrents. Let us demonstrate this by considering simple scheme of figure \ref{fig_balanced}, where the annihilation operator of the input quantum state is given by a strong mean field ($\alpha=|\alpha|e^{i\phi}$) added by quantum fluctuations ($\delta \hat{a}$).
\begin{figure}[ht]
    \centering
    \includegraphics[width=.20\textwidth]{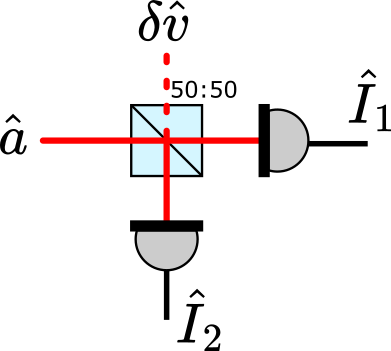}
    \caption{{Balanced detection scheme. The input field $\hat{a}$ is equally divided in a beamsplitter and directed to photodetectors. Vacuum fluctuations ($\delta\hat{v}$) are accounted in the open beamsplitter input. The output photocurrents are given in equations (\ref{eq_i1}) and (\ref{eq_i2}). }}
    \label{fig_balanced}
\end{figure}
After the $50:50$ splitting operation,
\begin{align}
    \hat{b}_1 &= \frac{1}{\sqrt{2}}\left(\alpha + \delta\hat{a}  + \delta \hat{v} \right) \\
    \hat{b}_2 &= \frac{1}{\sqrt{2}}\left(\alpha + \delta\hat{a}  - \delta \hat{v} \right)
\end{align} 
where $\hat{b}_1$ and $\hat{b}_2$ are corresponded to the two output arms of a beamsplitter and $\delta \hat{v}$ refers to vacuum fluctuations. The intensity on each of the arms are then given by
\begin{align}
    \label{eq_i1}
    \hat{I}_1 & =  \frac{1}{{2}}\left(|\alpha|^2 + |\alpha| \delta\hat{x}^{\phi}   + |\alpha|\delta \hat{v} \right) \\
    \label{eq_i2}
    \hat{I}_2 &= \frac{1}{{2}}\left(|\alpha|^2 + |\alpha| \delta\hat{x}^{\phi}   - |\alpha|\delta \hat{v} \right),
\end{align} 
where we disregarded terms that are quadratic on the fluctuations. The field quadrature $\hat{x}^{\phi} = \delta\hat{a} e^{-i\phi} + \delta\hat{a}^{\dagger} e^{i\phi}$ is defined by the relative phase between the carrier and the sidebands $\phi$, which we can vary with the aid of our analysis cavities. Separating the mean field (DC signal) from the fluctuations (AC signal), it is evident that the subtraction of the photocurrents will result in the vacuum fluctuations amplified by the amplitude of the field, which defines the shot noise. On the other hand, the sum carries the quadrature information, which can be directly compared to the shot noise level.}

\begin{figure}[ht]
    \centering
    \includegraphics[width=.98\textwidth]{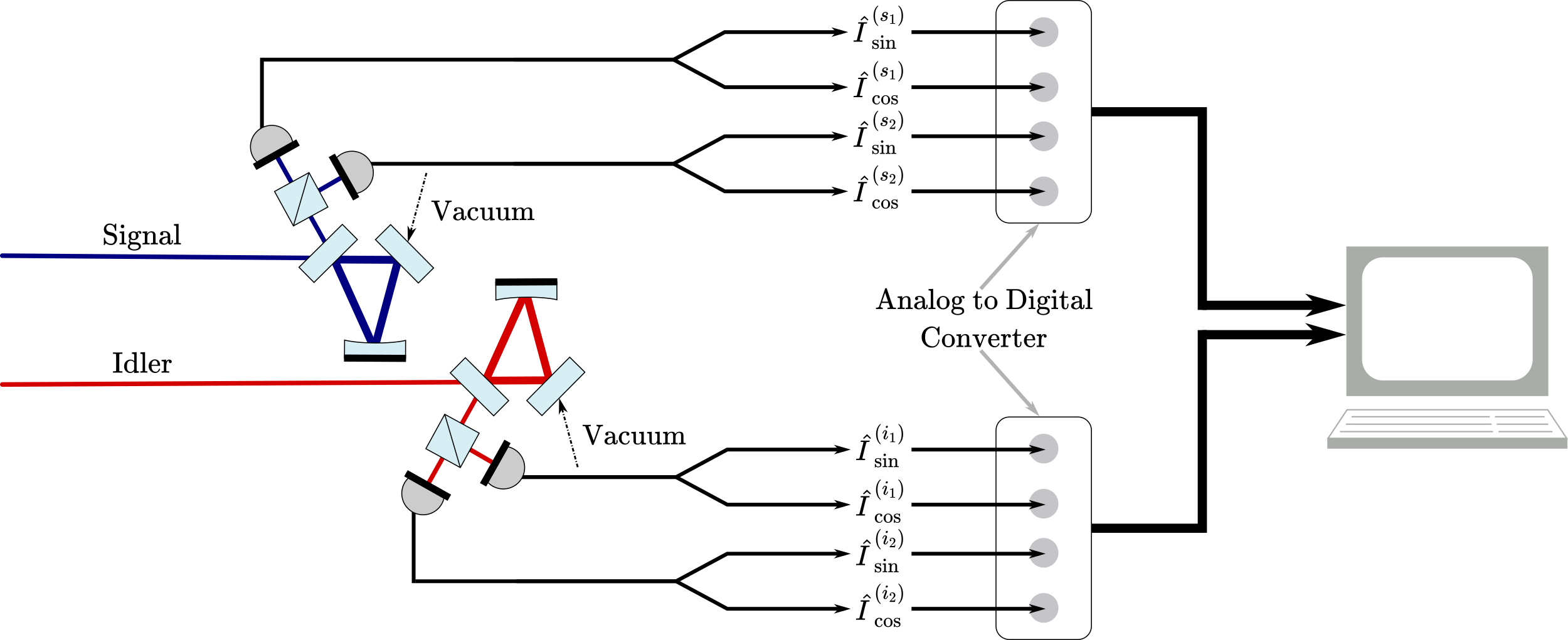}
    \caption{Detection setup scheme. After passing through the resonators, signal and idler undergoes a balanced detection. The double demodulated photocurrents, see figure \ref{fig_demod}, are individually acquired by analog to digital converters and computationally analyzed.}
    \label{fig_detectionscheme}
\end{figure}

Afterwards, each detected signal is sent to double demodulation, as described in section \ref{sec_resonatordetection}, and acquired in an analog-to-digital conversion system, figure \ref{fig_detectionscheme}. We then digitally process the different signals individually. Results for different measurements are shown in the main text.

Since we have access to all signals, we could characterized and digitally match the gain of the detector pairs of each balanced detection. After gain compensation, the expression for the normalized noise, with correction for the electronic noise, attained by a balanced detection with detectors labeled by $1$ and $2$ is given by
\begin{equation}
\label{eq_singlemeasurecorr}
    \Delta^2 \hat{x} = \frac{\Delta^2 \left( V_{\textrm{HF}}^{(1)} + V_{\textrm{HF}}^{(2)} \right) - \Delta^2 e^{(1)}_{} - \Delta^2 e^{(2)}_{}}{\Delta^2 \left( V_{\textrm{HF}}^{(1)} - V_{\textrm{HF}}^{(2)} \right) - \Delta^2 e^{(1)}_{} - \Delta^2 e^{(2)}_{}},
\end{equation}
where $V_{\textrm{HF}}^{(n)}$ is the high-frequency component of the photocurrent converted by a built-in transimpedance amplifier into a voltage signal, that if further demodulated (figure \ref{fig_demod}), and $\Delta^2 e^{(n)}$ is the electronic background noise intrinsic to the detection system. Thus we can obtain the  quadrature fluctuations $\Delta^2 \hat{x}$ normalized to the shot noise level. This equation was used to present the data shown in figure 2 (a) of the main text.

At last, it is straightforward to carry out the presented analysis in order to include the correlations of different fields, as in presented in figure 2 (b) of main text. The noise present in the sum and subtraction subspaces is then given by
\begin{align}
\label{eq_subspaces}
    \Delta^2 \hat{x}_{\pm}^{} &= \frac{1}{2} \left(\frac{\Delta^2 V^{(s_+)}_{\textrm{HF}} - \Delta^2 e^{(s)}_{}}{\Delta^2 V^{(s_-)}_{\textrm{HF}} - \Delta^2 e^{(s)}_{}} + \frac{\Delta^2 V^{(i_+)}_{\textrm{HF}} - \Delta^2 e^{(i)}_{}}{\Delta^2 V^{(i_-)}_{\textrm{HF}} - \Delta^2 e^{(i)}_{}} \right) \nonumber \\
    &\pm \frac{ \left\langle{V^{(s_+)}_{\textrm{HF}} V^{(i_+)}_{\textrm{HF}}}\right\rangle}{\sqrt{\left(\Delta^2 V^{(s_-)}_{\textrm{HF}} - \Delta^2 e^{(s)}_{}\right)\left(\Delta^2 V^{(i_-)}_{\textrm{HF}} - \Delta^2 e^{(i)}_{}\right)}},
\end{align}
where $V^{(s_\pm,i_\pm)}_{\textrm{HF}} = V^{(s_1,i_1)}_{\textrm{HF}} \pm V^{(s_2,i_2)}_{\textrm{HF}}$, and $e^{(s,i)} = e^{(s_1,i_1)} + e^{(s_2,i_2)}$.

\section{Measurement Example}

{
In this section we explicitly give the covariance matrix elements retrieved from the fittings shown in figures 2 and 3 of the main text. This example is referent to the measurement with the optimum squeezing level. Table \ref{table_parameters} shows the terms obtained by the individual spectral densities shown in figure 2 (a) of the main text. We determined the shown parameters by fitting equation (\ref{eq_spectraldensitycavity}) to our experimental data. }
\begin{table}[htbp]
\centering
\caption{Parameters determined by the power spectrum.}
\begin{tabular}{cccc}
\hline
\textbf{Parameter} & \textbf{Related Correlations}	&	\textbf{Mean Value} & \textbf{Standard Deviation} \\
\hline
$\alpha^{(s)}$ & $\Delta^2 \hat{p}^{(s)}_{\mathfrak{s}} = \Delta^2 \hat{q}^{(s)}_{\mathfrak{a}}$	& $10.44$ & $0.03$   \vspace{.1 cm} \\
$\beta^{(s)}$ & $\Delta^2 \hat{q}^{(s)}_{\mathfrak{s}} = \Delta^2 \hat{p}^{(s)}_{\mathfrak{a}}$ & $12.51$ & $0.09$ \vspace{.1 cm} \\
$\gamma^{(s)}$ & $\left\langle \hat{p}^{(s)}_{\mathfrak{s}}\hat{q}^{(s)}_{\mathfrak{s}}\right\rangle = - \left\langle\hat{p}^{(s)}_{\mathfrak{a}}\hat{q}^{(s)}_{\mathfrak{a}}\right\rangle$	& $-1.36$& $0.05$ \vspace{.1 cm} \\
$\delta^{(s)}$ & $\left\langle \hat{p}^{(s)}_{\mathfrak{s}}\hat{p}^{(s)}_{\mathfrak{a}} \right\rangle =  \left\langle \hat{q}^{(s)}_{\mathfrak{s}}\hat{q}^{(s)}_{\mathfrak{a}}\right\rangle$ & $-0.1$ & $0.3$ \vspace{.1 cm} \\
$\alpha^{(i)}$ & $\Delta^2 \hat{p}^{(i)}_{\mathfrak{s}} = \Delta^2 \hat{q}^{(i)}_{\mathfrak{a}}$	& $11.04$ & $0.04$   \vspace{.1 cm} \\
$\beta^{(i)}$ &  $\Delta^2 \hat{q}^{(i)}_{\mathfrak{s}} = \Delta^2 \hat{p}^{(i)}_{\mathfrak{a}}$   & $12.0$ & $0.1$ \vspace{.1 cm} \\
$\gamma^{(i)}$ & $\left\langle \hat{p}^{(i)}_{\mathfrak{s}}\hat{q}^{(i)}_{\mathfrak{s}}\right\rangle = - \left\langle \hat{p}^{(i)}_{\mathfrak{a}}\hat{q}^{(i)}_{\mathfrak{a}}\right\rangle$	& $-0.87$ & $0.06$ \vspace{.1 cm} \\
$\delta^{(i)}$ & $\left\langle\hat{p}^{(i)}_{\mathfrak{s}}\hat{p}^{(i)}_{\mathfrak{a}}\right\rangle =  \left\langle \hat{q}^{(i)}_{\mathfrak{s}}\hat{q}^{(i)}_{\mathfrak{a}}\right\rangle$  & $-0.7$ & $0.3$ \vspace{.1 cm} \\
\hline
\label{table_parameters}
\end{tabular}
\end{table}

{
Next, we present the remaining parameters, obtained by the fittings presented in figure 3 of the main text, in table \ref{table_crossparameters}. The computational method adopted simultaneously fit the $8$ presented curves, retrieving the $15$ of the $16$ parameters necessary for the full reconstruction of the density matrix. We note that the parameter $\mu$, related to intensity correlations, was fixed by the raw squeezing measurement, shown in figure 2 (b) of the main text.}
\begin{table}
\centering
\caption{Parameters determined by the cross-correlation functions.}
\begin{tabular}{cccc}
\hline
\textbf{Parameter}	& \textbf{Related Correlations} &	\textbf{Mean Value} & \textbf{Standard Deviation} \\
\hline 
$\mu$ & $\left\langle\hat{p}^{(s)}_{\mathfrak{s}} \hat{p}^{(i)}_{\mathfrak{s}}\right\rangle = \left\langle\hat{q}^{(s)}_{\mathfrak{a}} \hat{q}^{(i)}_{\mathfrak{a}}\right\rangle$ & $10.1$ & $0.2$  \vspace{.1 cm} \\
$\nu$ & $\left\langle\hat{q}^{(s)}_{\mathfrak{s}} \hat{q}^{(i)}_{\mathfrak{s}}\right\rangle = \left\langle\hat{p}^{(s)}_{\mathfrak{a}} \hat{p}^{(i)}_{\mathfrak{a}}\right\rangle$ & $0.57$ & $0.09$ \vspace{.1 cm} \\
$\kappa$ & $\left\langle\hat{p}^{(s)}_{\mathfrak{s}} \hat{p}^{(i)}_{\mathfrak{a}}\right\rangle = \left\langle\hat{q}^{(s)}_{\mathfrak{s}} \hat{q}^{(i)}_{\mathfrak{a}}\right\rangle$ & $-0.50$& $0.06$ \vspace{.1 cm} \\
$\lambda$ & $- \left\langle\hat{p}^{(a)}_{\mathfrak{s}} \hat{p}^{(i)}_{\mathfrak{s}}\right\rangle = - \left\langle\hat{q}^{(a)}_{\mathfrak{a}} \hat{q}^{(i)}_{\mathfrak{s}}\right\rangle$ & $1.84$ & $0.06$ \vspace{.1 cm} \\
$\xi$ & $\left\langle\hat{p}^{(s)}_{\mathfrak{s}} \hat{q}^{(i)}_{\mathfrak{s}}\right\rangle = - \left\langle\hat{q}^{(s)}_{\mathfrak{a}} \hat{p}^{(i)}_{\mathfrak{a}}\right\rangle$ & $-1.45$ & $0.06$ \vspace{.1 cm} \\
$\zeta$ & $\left\langle\hat{q}^{(s)}_{\mathfrak{s}} \hat{p}^{(i)}_{\mathfrak{s}}\right\rangle = \left\langle\hat{p}^{(s)}_{\mathfrak{a}} \hat{q}^{(i)}_{\mathfrak{a}}\right\rangle$ & $-0.74$ & $0.06$ \vspace{.1 cm} \\
$\eta$ & $- \left\langle\hat{p}^{(s)}_{\mathfrak{s}} \hat{q}^{(i)}_{\mathfrak{a}}\right\rangle = \left\langle\hat{q}^{(s)}_{\mathfrak{a}} \hat{p}^{(i)}_{\mathfrak{s}}\right\rangle$ & $-0.66$ & $0.02$ \vspace{.1 cm} \\
$\tau$ & $- \left\langle\hat{q}^{(s)}_{\mathfrak{s}} \hat{p}^{(i)}_{\mathfrak{a}}\right\rangle = \left\langle\hat{p}^{(s)}_{\mathfrak{a}} \hat{q}^{(i)}_{\mathfrak{s}}\right\rangle$ & $-2.62$ & $0.09$ \vspace{.1 cm} \\
\hline
\label{table_crossparameters}
\end{tabular}
\end{table}

{
The presented parameters completely define the covariance matrix presented in equations (\ref{eq_covmatrixred})--(\ref{eq_covmatcorr}). This procedure was done to reconstruct $22$ different states, generated with different pump powers. Once with the covariance matrices in hand, we computed their physicality and purity, as described in the following section.
}

\section{Physicality, Purity and Entanglement Criterion}

The covariance matrix gives us a complete description of a Gaussian state, hence, such states are vastly explored with the covariance matrix formalism
\cite{serafini2017quantum}. First, we guaranteed the validity of the reconstructed state, that is, if our tomography retrieved a physical state. For $\mathbb{V}^{}_{}$ to be a valid representation of a physical density matrix, the uncertainty relation
\begin{equation}
\label{eq_symplecticuncertainty}
    \mathbb{V}^{}_{} + i\mathbb{W} \geq 0,\qquad \mbox{with}\qquad 
\mathbb{W}=\bigoplus_{i=1}^N\begin{bmatrix}
    0 & 1 \\
    -1 & 0.
\end{bmatrix}
\end{equation}
namely the Robertson-Schrödinger uncertainty relation, must be fulfilled. Alternatively, by Williamson's theorem \cite{williamson1936algebraic}, any $n$-mode Gaussian state can be represented in a diagonal form undergoing a transformation
\begin{equation}
\label{eq_symplectieigenvals}
    \mathbb{V}^{}_D = S \mathbb{V}^{}_{} S^{\textrm{T}} = \textrm{Diag} \{ \nu_1, \nu_1, \nu_2, \nu_2, \ldots, \nu_n, \nu_n \}
\end{equation}
where $S$ and $S^{\textrm{T}}$ are symplectic operations. The uncertainty relation of equation (\ref{eq_symplecticuncertainty}) holds in this representation since $\mathbb{W}$ is invariant under symplectic transformations. Therefore, the eigenvalues of $\mathbb{V}^{}_D$ must respect the condition $\nu_j \geq 1, j = \{1,2,\ldots,n\}$. A practical way to determine the symplectic eigenvalues is by diagonalizing the matrix $(\mathbb{V}\mathbb{W})^2$ whose eigenvalues are given by $(-\nu_j^2), j = \{1,2,\ldots,n\}$ \cite{adesso2014continuous}.

The purity of the state, that is, how close the system is to a pure state, is given by \cite{paris2003purity}
\begin{equation}
    \mathfrak{p} = \frac{1}{\sqrt{\textrm{Det}(\mathbb{V}^{}_{})}} = \frac{1}{\Pi_{j=0}^{n} \nu_j^2}.
    \label{eq_purity}
\end{equation}
{A pure state is indicated by $\mathfrak{p} = 1$.} This equation was used to calculate the points shown in figure 5 of the main text. {There, one can see the degradation of the purity as the system quadratures become uncorrelated. This can also be seen in the squeezing degradation and the increasing phase noise sum, respectively shown in figures 3(c) and 3(d) of the main text.}

Finally, we analyzed an entanglement criterion for the reconstructed states. A Gaussian state is separable if and only if there exists the covariance matrices $\mathbb{V}_A$ and $\mathbb{V}_B$ for the respective subsystems $A$ and $B$ such that they respect the inequality \cite{werner2001bound}
\begin{equation}
    \mathbb{V} \geq \mathbb{V}_A \oplus \mathbb{V}_B.
\end{equation}
Although general for Gaussian states, this criterion is not very useful in practice. Restricting the number of degrees of freedom of the system, more applicable criteria can be used. 

\begin{figure}[ht]
    \centering
    \includegraphics[width=0.65\textwidth]{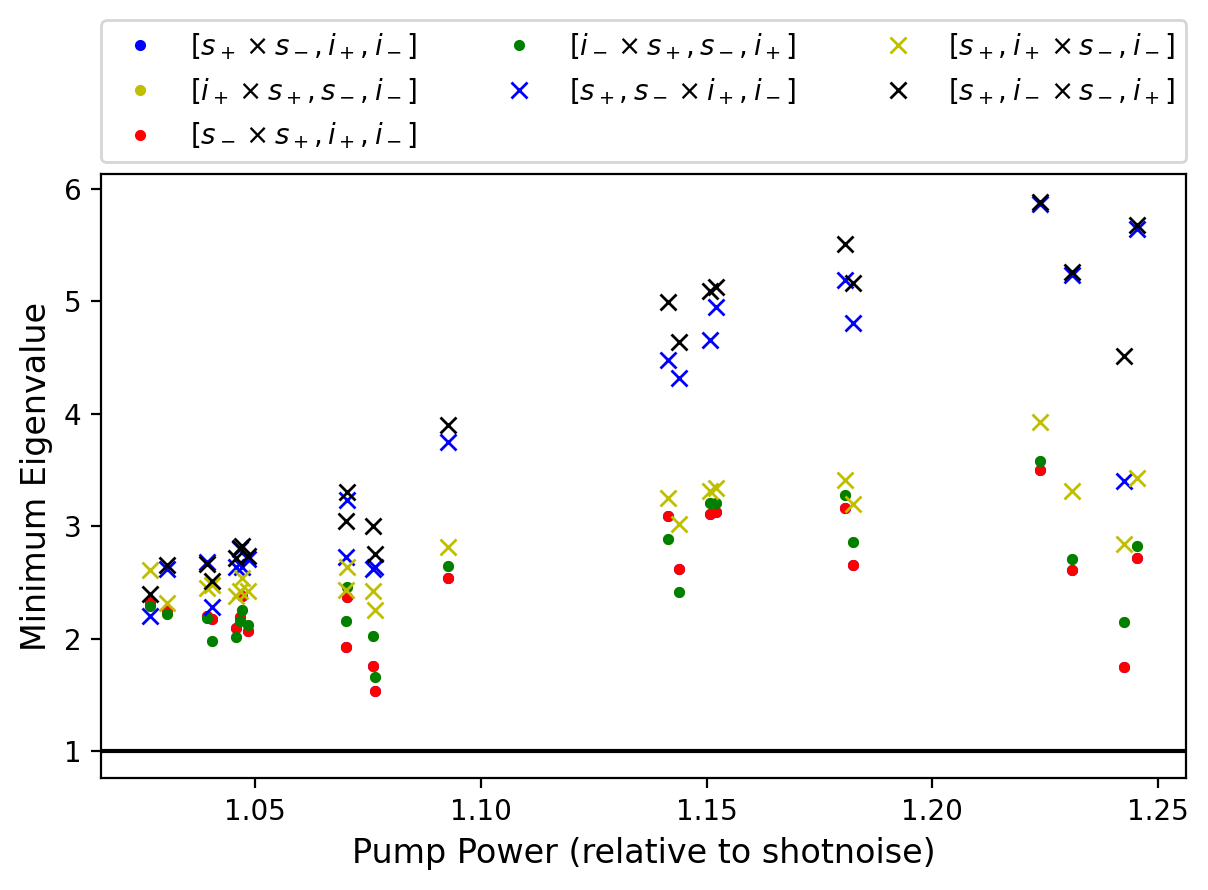}
        \caption{PPT test for different transpositions for the reconstructed covariance matrices. The dashed line indicating the value $1$ represents the condition of physicality. None of the substates are entangled since no minimum symplectic eigenvalue is below $1$.}
    \label{fig_PPTtest}
\end{figure}

A practical approach is done by the analysis of the partial transposition of $\mathbb{V}$ with a method known as positive partial transpose (PPT) criterion \cite{simon2000peres}. The partial transposition of a quantum state of $(m+n)$ modes with respect to the $n$ partition is calculated as
\begin{equation}
    \tilde{\mathbb{V}} = T \mathbb{V} T,
\end{equation}
with
\begin{equation}
    T = \mathbb{1}_{2m} \oplus \Sigma_n, \; \; \; \; \Sigma_n = \bigoplus_{j=1}^{n} \sigma_z^{},
\end{equation}
where $\mathbb{1}_{2m}$ is the identity matrix in a $2m$ space $\sigma_z$ is the Pauli-z matrix. That is, if the physicality criterion is violated after partial transposition, the transposed subsystem is entangled with the rest of the system. We verified the minimum symplectic eigenvalues for the different transpositions of all reconstructed covariance matrices. The results are compiled in figure \ref{fig_PPTtest}, where no evidence of entanglement is present.


\section{Phase Self-Modulation}

\begin{figure}[ht]
    \centering
    \includegraphics[width=0.55\textwidth]{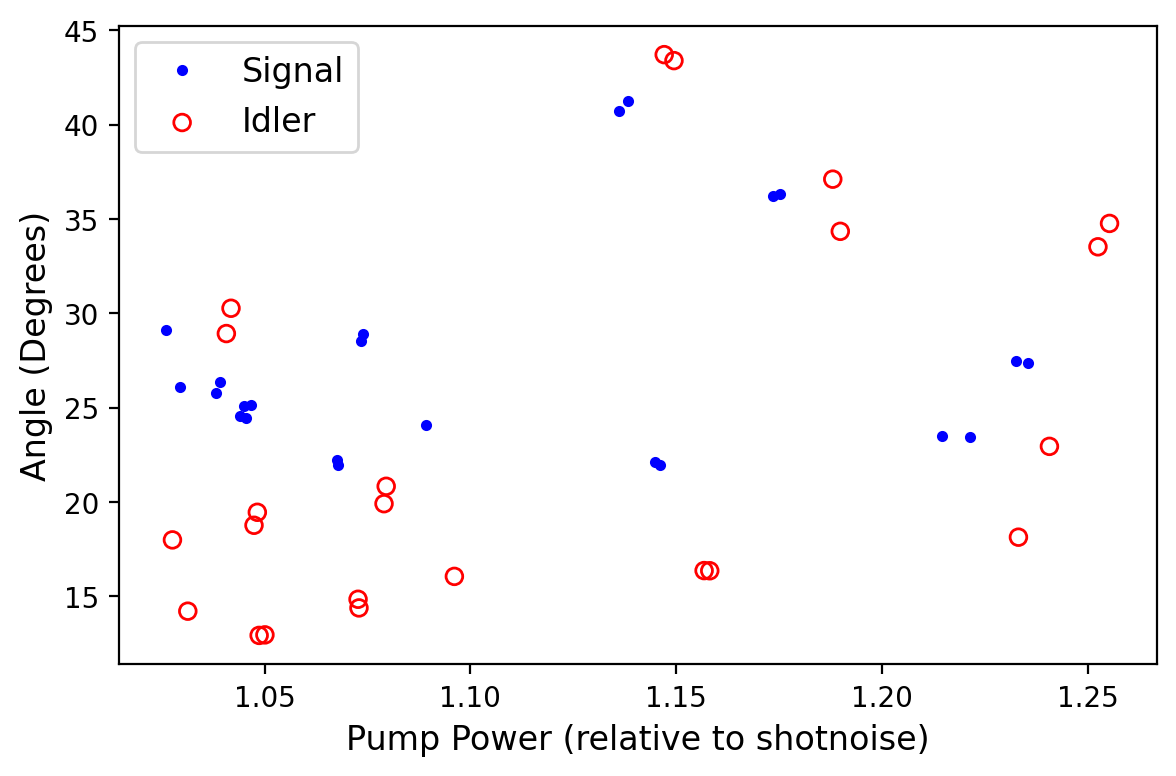}
        \caption{Rotation angles that independently aligns the axis of signal and idler noise ellipses. All the measurements present appreciable angles, indicating the influence of phase modulations in the dynamics of the system.}
    \label{fig_rotationangles}
\end{figure}

\begin{figure}[ht]
    \centering
    \includegraphics[width=0.85\textwidth]{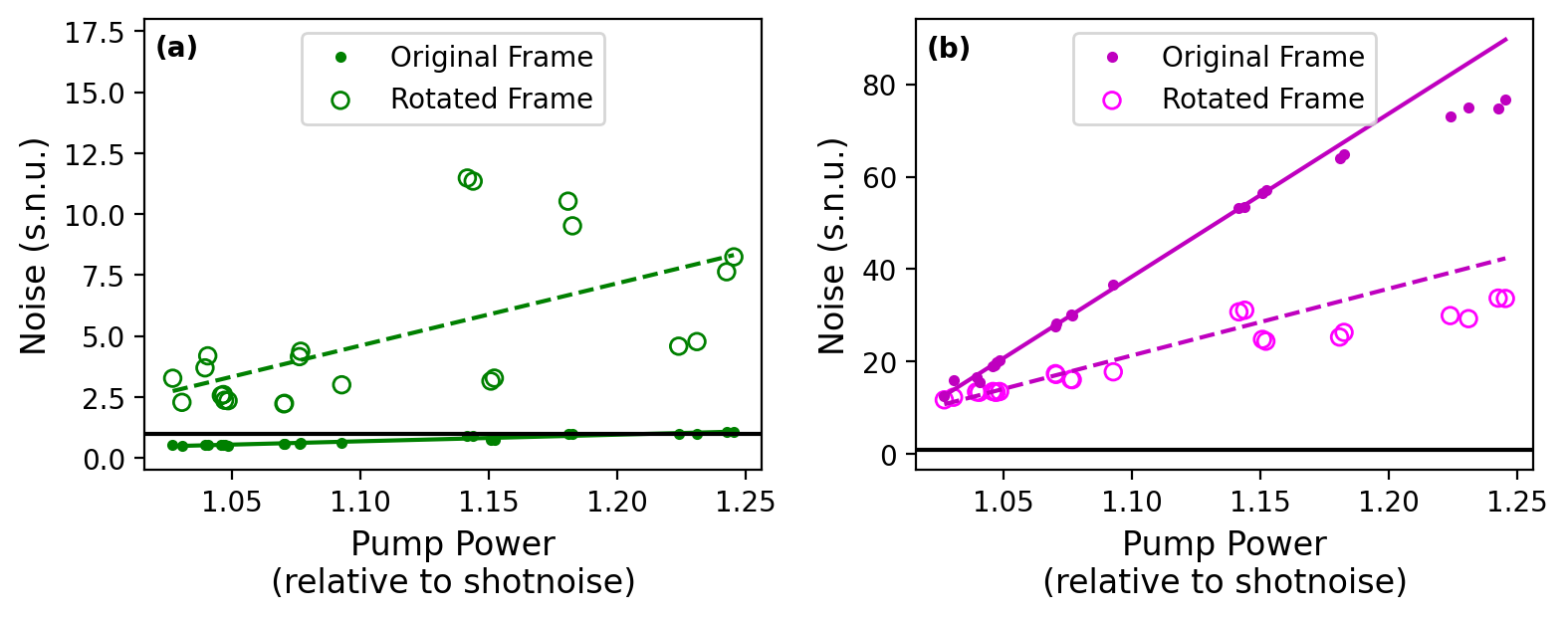}
    \caption{Comparison of the amplitude subtraction (a) and phase sum (b) noises between the original and rotated frames. The amplitude squeezing is completely lost in the new frame.}
    \label{fig_compsumsubspaces}
\end{figure}

By taking a spectral matrix approximation \cite{barbosa2013quantum}, where we return to the two-mode approximation, correlations between phase and amplitude quadratures are observed. This is equivalent to distortions in the noise ellipse in the phase space, a predicted consequence in $\chi^{(3)}$ systems due to  Kerr-effect phase modulations \cite{ferrini2014symplectic,gonzalez2017third}. We then perform a frame rotation to align the quadratures with the main axis of the noise ellipse hoping to increase the observed correlations \cite{gonzalez2017third}. Although relevant rotation angles were observed for all reconstructed states, figure \ref{fig_rotationangles}, no enhancement on the correlations were observed.
In fact, we lose the non-classical characteristics of our measurements when looking in this new aligned frame, figure \ref{fig_compsumsubspaces} (b). We attribute this effect to the contamination of excessive phase noise in the amplitude quadrature through cross-phase modulation effect \cite{gonzalez2017third}. On the other hand, phase sum noise was drastically decreased, as shown in figure \ref{fig_compsumsubspaces} (b).

\newpage

\bibliography{sample}
